\documentclass[useAMS,usenatbib]{mn2e}
\usepackage{aas_macros}
\usepackage{graphicx}
\usepackage{amssymb}
\usepackage{amsmath}
\def\psad{$P^2SAD$}
\usepackage{color}

\title[Clustering in the Phase Space of Dark Matter Haloes. I. Results from the Aquarius simulations]
{Clustering in the Phase Space of Dark Matter Haloes. I. Results from the Aquarius simulations}
\author[Jes\'us Zavala, Niayesh Afshordi]{\parbox{18cm}{Jes\'us
    Zavala$^{1,2,3}$\thanks{e-mail: jzavalaf@uwaterloo.ca} and Niayesh Afshordi$^{1,2}$\vspace{0.3cm}}\\ 
$^{1}$Perimeter Institute for Theoretical Physics, 31 Caroline St. N., Waterloo, ON, N2L 2Y5, Canada\\
$^{2}$Department of Physics and Astronomy, University of Waterloo, Waterloo, Ontario, N2L 3G1, Canada\\
$^{3}$Canadian Institute for Theoretical Astrophysics, University of Toronto, Toronto, Ontario M5S 3H8, Canada}

\begin{document}



\maketitle

\label{firstpage}

\begin{abstract}

We present a novel perspective on the clustering of dark matter in phase space by defining the
particle phase space average density (\psad) as a two-dimensional extension of the two-point
correlation function averaged within a certain volume in phase space. This statistics is a sensitive 
measure of small scale (sub-)structure of dark matter haloes. 
By analysing the structure
of \psad~in Milky-Way-size haloes using the 
Aquarius simulations, we find it to be 
nearly universal at small scales, i.e. small separations in phase space, 
where substructures dominate. This remarkable universality
occurs across time and in regions of substantially different ambient densities (by nearly four orders of magnitude), with typical
variations in \psad~of a factor of a few. The maximum variations occur in regions where 
substructures have been strongly disrupted. 
The universality is also preserved across haloes of similar mass but diverse mass accretion histories
and subhalo distributions.
The universality is also broken at large scales, 
where the smooth dark matter distribution in the
halo dominates. Although at small
scales the structure of \psad~is roughly described by a subhalo model, we argue that the simulation data
is better fitted by a family of superellipse contours. This functional shape is inspired by a model that
extends the stable clustering hypothesis into phase space. In a companion paper, we refine this model and show
its advantages as a method to obtain predictions for non-gravitational signatures of dark matter.

\end{abstract}

\begin{keywords}
cosmology: dark matter - methods: N-body simulations.
\end{keywords}

\section{Introduction}

The gravitational clustering of dark matter has been studied intensely over the last few decades, mostly in the context of our
understanding of how galaxies form and evolve within dark matter haloes. With the increase in computational power and improvement
in numerical techniques, $N-$body simulations have given us a detailed picture of how dark matter clusters in space from 
large cosmological scales ($\sim$~Gpc) down to the inner parts of galactic haloes where baryons clearly dominate the gravitational potential
($\lesssim100$~pc). Although such progress has led to a consensus on a variety of attributes that result from the clustering of dark
matter into haloes, e.g. the universality of the density profiles \citep[originally established in][]{Navarro_96,Navarro_97}, there are 
open issues at the scales 
that are unresolved in current simulations, such as the abundance, distribution and internal structure of dark matter haloes in this regime.

The clustering of dark matter at sub-kpc scales is of key importance for predictions for the hypothetical non-gravitational signatures
of dark matter. Since the resolution of current simulations
is many orders of magnitude above the cutoff mass of bound haloes for the majority of dark matter candidates, it is necessary to
rely on uncertain extrapolations to the unresolved scales. Such uncertainty impacts the predicted experimental signals for both 
dark matter annihilation (indirect detection) and dark matter searches in laboratories (direct detection).   

Although most studies on the distribution of dark matter have focused on spatial clustering, a more complete picture emerges by analysing 
its phase space structure \citep[e.g.][]{Bertschinger_85,Vogelsberger_08,Diemand_08,Afshordi_09,Vogelsberger_09,Vogelsberger_11,Maciejewski_11}. 
This is 
particularly important for dark matter detection efforts if the microscopic interaction of dark matter particles
is velocity dependent \citep[examples are the so-called Sommerfeld-enhanced models, e.g.][]{Arkani_Hamed_09}. 
In a series of two papers, we study the clustering of dark matter in the phase space of haloes using the
simulation suite from the Aquarius project \citep{Springel_08}. We do so by defining a novel quantity: the particle phase space average 
density (\psad), 
which is an extension in phase space of the concept of the two-point correlation function (2PCF). In the present paper, we introduce this
quantity and show, remarkably, that, at small scales (i.e. small separations in phase space), is nearly universal in time and across diversely assembled haloes, while also showing 
relatively small variations across spatial regions with substantially different densities. 

It is important to note that \psad~is basically a measure of the coarse-grained phase space density (averaged around particles), as a function 
of coarse-graining scale. Naturally, our measurements are thus limited by the dark matter clustering resolved in the simulation. 
The numerically resolved coarse-graining scale is many orders of magnitude
above that of the fine-grained phase space density which captures the primordial dark matter streams, and stretches and folds in phase 
space through the growth of structure \citep[e.g.,][]{Vogelsberger_08}. Therefore, our measurement of \psad~is not sensitive to 
these very small clustering scales. It is
also different from the commonly used pseudo-phase-space density defined from halo-centric shell averages of the density
$\rho$ and velocity dispersion $\sigma_{\rm vel}$:  $Q\equiv\rho/\sigma_{\rm vel}^3$ \citep[e.g.][]{Taylor_01}.

In a companion paper \citep{Zavala_13} we present
a model of the \psad~inspired by an extension into phase space of the stable clustering hypothesis originally introduced
by \citet{Davis_77}, and recently presented in \citet{Afshordi_10} to predict the phase space clustering of gravitationally 
bound dark matter substructure. We refine this model by incorporating a simplified prescription of the effects of tidal disruption. 
Finally, we present some interesting implications for dark matter detection efforts by extrapolating this physically motivated model to the small
unresolved scales.

This paper is organised as follows. In Section \ref{2PCF} we define \psad~and test the algorithm we use to compute it in a simulation. In Section
\ref{section_3} we investigate the behaviour of \psad~in the smooth halo (Section \ref{smooth}), and in substructures and across time (Section 
\ref{evolution}). In Section \ref{fitting} we provide fitting functions for \psad~in the regime where subhaloes dominate. We study the variations
of \psad~across regions of different ambient density and across haloes with different merger histories in Sections \ref{vol_sec} and \ref{mahs}, 
respectively. Finally, a summary and our main conclusions are given in section \ref{conclusions}.

\section{Particle Phase Space Average Density (\psad)}\label{2PCF}

The 2PCF between two points in phase space separated by ${\bf \Delta x}$ and ${\bf \Delta v}$ is given in terms of the phase space distribution:
\begin{eqnarray}\label{2pcf_eq_0}
\langle f({\bf x},{\bf v})f({\bf x}+{\bf \Delta x},{\bf v}+{\bf \Delta v})\rangle_{{\cal V}_6}\nonumber\\
\equiv \frac{1}{{\cal V}_6}\int_{{\cal V}_6} d^3{\bf x}d^3{\bf v} f({\bf x},{\bf v})f({\bf x}+{\bf\Delta x},{\bf v}+{\bf \Delta v})\nonumber\\
\end{eqnarray}
where we average over a volume ${\cal V}_6$ of the phase space. For convenience, we study the 2PCF in this paper by analyzing instead 
the mass-weighted average phase space density of dark matter, lying at spheres of radius 
$\Delta x=\vert{\bf \Delta x}\vert$ and $\Delta v=\vert{\bf \Delta v}\vert$, 
in position and velocity spaces respectively:
\begin{eqnarray}\label{2pcf_eq}
\Xi(\Delta x, \Delta v)\equiv
\frac{\int_{{\cal V}_6}d^3{\bf x}d^3{\bf v} f({\bf x},{\bf v}){\tilde f}({\bf x}, {\bf v}; \Delta x,\Delta v)}
{\int_{{\cal V}_6} d^3{\bf x}d^3{\bf v} f({\bf x},{\bf v})}
\end{eqnarray}
where ${\tilde f}$ is the average of $f$ over solid angles in ${\bf\Delta x}$ and ${\bf\Delta v}$:
\begin{equation}
  {\tilde f}({\bf x}, {\bf v}; \Delta x,\Delta v) = \frac{ \int d\Omega_{{\bf \Delta x}} d\Omega_{{\bf \Delta v}} f({\bf x}+{\bf\Delta x},{\bf v}+{\bf \Delta v}) } {\int d\Omega_{{\bf \Delta x}} d\Omega_{{\bf \Delta v}}}.
\end{equation}
In this form, $\Xi(\Delta x, \Delta v)$ is the auto-correlation function of the fine-grained phase space density. 
Since we deal with coarse-grained quantities from here on, and for reasons that become more clear in the next subsection, we call 
$\Xi(\Delta x, \Delta v)$ Particle Phase Space Average Density, or in short: \psad.
\subsection{Numerical algorithm and testing}\label{l_estimator}

\begin{figure}
\center{
\includegraphics[height=4.5cm,width=8.0cm]{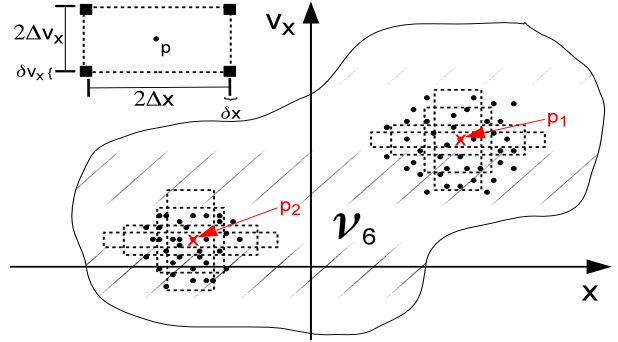}
}
\caption{Sketch illustrating how the the particle phase space average density (\psad) is estimated in a simulation in the
one-dimensional case. For each particle
$p$ in a random sample of particles in a region of volume ${\cal V}_6$, we count the number of particles $N_p$ that lie at a separation in phase space 
in the corners of the rectangles of sizes $(\Delta x-\delta x/2, \Delta v_x-\delta v_x/2)$ and $(\Delta x+\delta x/2, \Delta v_x+\delta v_x/2)$ centred in $p$. 
The \psad~(Eq.~\ref{2pcf_eq_sim}) is then defined as the average phase space density of particles in these rectangular corners of size $(\delta x,\delta v_x)$.
The \psad~is then a measure of average clustering, at the $(\Delta x, \Delta v)$ scale, in the phase space of the region of volume ${\cal V}_6$.} 
\label{Fig_diagram} 
\end{figure}

In a simulation, the dark matter is represented by a discrete set of $N$ particles, where each has a mass $m_p$ many orders of magnitude
above the free-streaming mass of cold dark matter (CDM) particles. In this discrete representation of the dark matter density field, 
Eq.~(\ref{2pcf_eq}) is estimated as:
\begin{equation}\label{2pcf_eq_sim}
\Xi(\Delta x, \Delta v)_{\rm sim}=\frac{m_p \langle N_p(\Delta x, \Delta v)\rangle_{{\cal V}_6}}{V_6(\Delta x,\Delta v)},
\end{equation}
where $\langle N_p\rangle$ is the average number of simulation particles within shells of thickness $(\delta x,\delta v)$ at a radius 
$\Delta x$ and $\Delta v$ in phase space, centred on each of the particles in the phase space volume ${\cal V}_6$ (see Fig.\ref{Fig_diagram}), and 
$V_6(\Delta x,\Delta v)$ is the phase space volume of a given shell. In the 
limit that the mass resolution of the simulation, $m_p \rightarrow 0$, $\Xi_{\rm sim}$ approaches $\Xi$ in 
Eq.~(\ref{2pcf_eq}). This is also why we use the name Particle Phase Space Average Density, or  \psad, since this quantity
describes the mean phase space density that a typical particle in volume ${\cal V}_6$ sees at a certain separation from its position
in phase space.

To compute the \psad~in an optimal way, we use an algorithm that selects a random sample $N_s$ of the total particles within
a given region of the simulation and computes $\langle N_p\rangle$ within this sample. 
Since the relative sampling error roughly scales as
$\sim1/\sqrt{N_s}$, a fixed value for the sample of particles
gives roughly the same ``noise'' level in the estimate of $\Xi_{\rm sim}$, independent of the resolution of the simulation. The fraction of
particles in the sample for a simulation with a total of $N_{\rm part}$ particles is then selected as $f_s=\alpha\times(10^7/N_{\rm part})$.
The sample is selected in the following way: for each of the $N_{\rm part}$ particles we randomly draw a number between 0 and 1 and
choose the particle if this number is $\leq f_s$. We have found that a value of $\alpha=0.6$ provides negligible errors (in most of the
phase space range explored in this paper) for a reasonable amount of computation time (see also Appendix). Unless otherwise stated, we use this 
sample size ($N_s=6\times10^6$) for the rest of this paper.

\begin{figure}
\center{
\includegraphics[height=8.0cm,width=8.0cm]{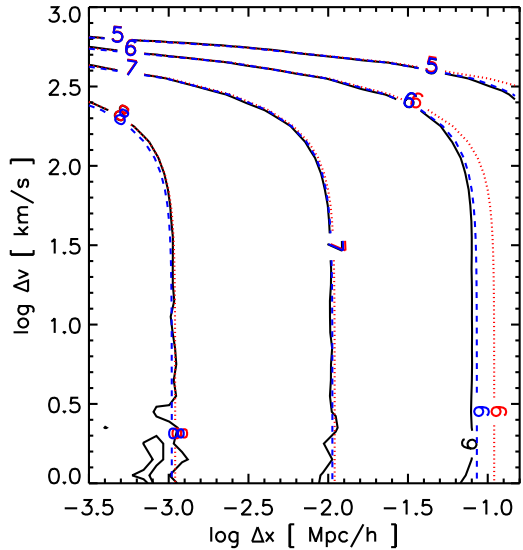}
}
\caption{Contours of the logarithm of the particle phase space average density (\psad) for a singular isothermal sphere with a cutoff radius of
$R_c=222.7$~kpc. The analytical result for the case $\Delta x\ll R_c$ is shown with a dotted red line while 
the result of evaluating Eq.~(\ref{2pcf_eq}) numerically is shown with a dashed blue line. The result of applying the full numerical code for
a Monte Carlo realization with $6\times10^6$ particles is shown in black. Poisson noise starts to dominate at the 
lowest separations in velocity, particularly at small values of $\Delta x$.} 
\label{Fig_MB} 
\end{figure}

To test the algorithm, we generate a Monte Carlo realization of $N_{\rm part}$ particles following a spherically symmetric distribution with a
density profile corresponding to a singular  isothermal sphere with a cutoff at a radius $R_c=222.7~{\rm kpc}$ ($M=1.37\times10^{12}$M$_\odot$). The 
corresponding phase space profile follows the Maxwell-Boltzmann distribution:
\begin{eqnarray}\label{MB}
  f({\bf x}, {\bf v})&=&\frac{A}{(2\pi\sigma_{\rm disp}^2)^{3/2}}{\rm exp}\left[(\phi(r)-v^2/2)/\sigma_{\rm disp}^2\right],\nonumber\\
  \phi(r)&=&2\sigma_{\rm disp}^2\left[1+{\rm ln}(R_c/r)\right]; \ \ \ \ \ r\leq R_c
\end{eqnarray}
where $A=\sigma_{\rm disp}^2/(2\pi {\rm G} e^2R_c^2)$, with $\sigma_{\rm disp}$ being the velocity dispersion, $v$ is the magnitude  
of the velocity vector, and we have used the fact that since the potential is central, it only depends on the radial distance $r$.
We note that for $r>R_c$, $f=0$ in Eq.~\ref{MB}.

If $\Delta x\ll R_c$ and we assume no cutoff in the velocity distribution, then Eq.~\ref{2pcf_eq} can be solved analytically:
\begin{equation}\label{MB_A}
  \Xi(\Delta x, \Delta v)_{\rm MB}=\frac{1}{64\pi^{1/2}G\sigma_{\rm disp} R_c\Delta x}{\rm exp}\left(\frac{-\Delta v^2}{4\sigma_{\rm disp}^2}\right)
\end{equation}
Recall that $\Delta x$ and $\Delta v$ are the radii in position and velocity space, respectively.

In Fig.~\ref{Fig_MB}, we show a comparison between the \psad~
computed from the Monte Carlo realization using our algorithm (black) and from Eq.~(\ref{MB_A}) (dotted red) 
With the exception of the region where the analytical result breaks down (i.e. $\Delta x\sim R$), the agreement is very good. 
The correct result is recovered when Eq.~(\ref{2pcf_eq}) is solved numerically (dashed blue) for any value of $(\Delta x, \Delta v$).
We note that for the thickness of the shells in phase space we have used logarithmic bins equally spaced with $\delta_{\rm log x}\sim0.1$ 
and $\delta_{\rm log v}\sim0.1$ throughout this paper (we have checked that our results are not changed appreciably by decreasing these values).

\section{$\Lambda$CDM haloes}\label{section_3}

To compute the \psad~in galactic $\Lambda$CDM haloes, we use the set of Aquarius haloes \citep{Springel_08}, which were simulated in the 
context of a WMAP1 cosmology and are examples of the type of Milky-Way-size haloes predicted by CDM. The Aquarius simulations are 
``zoom simulations'' from a larger box
parent simulation where a small region containing the volume of interest at $z=0$, i.e., the MW-size halo up to a few times its virial radius,
is re-sampled and re-simulated with a higher number of lower mass particles. The remaining volume is sampled with decreasing mass resolution, 
but large enough to accurately represent the long-range tidal field. In this paper we consider only
the particles within the high resolution region and mainly focus on those belonging to the virialized region of the main halo 
(defined as gravitationally self-bound objects by the structure finder SUBFIND, \citealt{Springel_01}). 
We note that the main halo is defined through a Friends-of-Friends method and thus it has a triaxial shape.
\psad~is clearly a function of location within the halo since its value depends on the phase space volume where the
average is done, ${\cal V}_6$ (see Eqs.~\ref{2pcf_eq}-\ref{2pcf_eq_sim}). While most of our results refer to \psad~averaged over the whole virialized 
region of the main halo, we discuss the environmental dependence of \psad~in Section \ref{vol_sec}.

\begin{table}
\begin{center}
\begin{tabular}{lllll}
\hline
Name & $m_{\rm p} [\mathrm{M}_\odot]$  & $\epsilon [\mathrm{pc}]$ & $M_{\rm 200} [\mathrm{M}_\odot]$ & $r_{\rm 200} [\mathrm{kpc}]$ \\
\hline
Aq-A-2   &  $1.370\times 10^4$  & 65.8  &   $1.842\times 10^{12}$ &  245.88 \\
Aq-A-3   &  $4.911\times 10^4$  & 120.5  &   $1.836\times 10^{12}$ &  245.64 \\
Aq-A-4   &  $3.929\times 10^5$  & 342.5  &   $1.838\times 10^{12}$ &  245.70 \\
Aq-A-5   &  $3.143\times 10^6$  & 684.9  &   $1.853\times 10^{12}$ &  246.37 \\
\hline
Aq-B-4   &  $2.242\times 10^5$  & 342.5  &   $8.345\times 10^{11}$ &  188.85 \\
\hline
Aq-C-4   &  $3.213\times 10^5$  & 342.5  &   $1.793\times 10^{12}$ &  243.68 \\
\hline
Aq-D-4   &  $2.677\times 10^5$  & 342.5  &   $1.791\times 10^{12}$ &  243.60 \\
\hline
Aq-E-4   &  $2.604\times 10^5$  & 342.5 &    $1.208\times 10^{12}$ &  213.63 \\
\hline
\end{tabular}
\end{center}
\caption{Parameters of the Aquarius haloes used in this paper: $m_{\rm p}$ is the particle
mass, $\epsilon$ is the  Plummer equivalent gravitational softening length,
$r_{\rm 200}$ and $M_{\rm 200}$ are, respectively, the virial radius and mass of the halo 
(defined as the mass enclosed in a sphere with mean density 200 times the critical value).}
\label{table_sims} \end{table}

In Table \ref{table_sims}, we list the main parameters of the set of Aquarius haloes we analyse in this paper. We 
focus mainly on the Aquarius-A halo that was simulated in 5 levels of increasing mass and spatial resolution. In particular
most of our results refer to Aq-A-2 and use the lower resolution levels for convergence tests\footnote{We note that we did not
use the highest resolution level, Aq-A-1, since it was deemed as computationally too expensive for the computation of \psad.}.
The different level-4 haloes are used to analyse diverse mass aggregation histories.

\begin{figure}
\includegraphics[height=8.0cm,width=8.0cm, angle=-90, trim=1.0cm 0.5cm 1cm 2.5cm, clip=true]{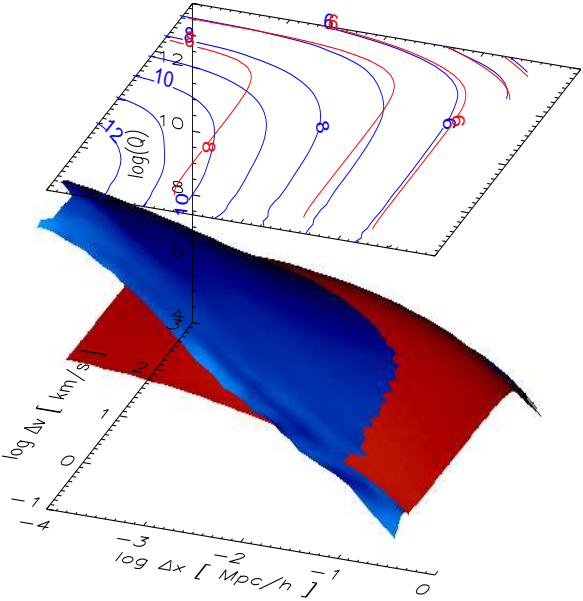} 
\caption{Surface and contour plot of the logarithm of the particle phase space average density (\psad) for the Aquarius A-halo resolution level 2 at $z=0$.
The average is done over the entire halo, including its substructure (blue) and for the smooth component of the host only (red). While
the former is estimated directly from the simulation data using Eq. (\ref{2pcf_eq_sim}), the latter is computed analytically using an Einasto density 
profile with a Maxwellian velocity distribution in Eq. (\ref{2pcf_eq}).} 
\label{Fig_host} 
\end{figure}

\subsection{Smooth main halo component}\label{smooth}

Fig.~\ref{Fig_host} shows a surface plot of log(\psad) computed with our algorithm for the Aq-A-2 halo considering
particles within the main halo and its subhaloes (blue). A 2D projection showing contours for log(\psad) is also shown on top.
At large separations in $(\Delta x,\Delta v)$, \psad~is dominated by the smooth dark matter distribution of the main halo shown in red
in Fig.~\ref{Fig_host}. To compute the latter, we used an Einasto density profile:
\begin{equation}\label{einasto}
\rho(r)=\rho_{-2}{\rm exp}\left(\frac{-2}{\alpha_e}\left[\left(\frac{r}{r_{-2}}\right)^{\alpha_e}-1\right]\right)
\end{equation}
where $\rho_{-2}$ and $r_{-2}$ are the density and radius at the point where the logarithmic density slope is -2, and $\alpha_e$ is the Einasto
shape parameter. We used the values of these
parameters from the fit to the Aq-A-2 halo given in \citet{Navarro_10}: $\alpha_e=0.163$, $\rho_{-2}=3.9\times10^{15}~{\rm M}_\odot/{\rm Mpc}^3$,
$r_{-2}=15.27~{\rm kpc}$. Furthermore, we assume that the velocity distribution of the smooth component is
Maxwellian with no cutoff in velocity and with a radially dependent 
1D velocity dispersion as given by the spherically averaged
pseudo-phase space density $\rho/\sigma_{\rm disp}^3\propto r^{-\chi}$ with $\chi=-1.917$ (see Table 2 of \citealt{Navarro_10}, 
and discussion therein). The normalization of $\sigma_{\rm disp}$ is given by its peak value: $\sigma_{\rm max}\sim V_{\rm max}/\sqrt{3}$, 
where $V_{\rm max}=208.49~{\rm km/s}$ is the maximum rotational velocity of the halo. Having assumed the density profile and the velocity 
distribution, we compute \psad~by numerically integrating Eq.~(\ref{2pcf_eq}).

There is a good match between the smooth halo modelling described before and the simulation results at large $(\Delta x,\Delta v)$. At
close separations though, it is clear that the contribution of the smooth component is subdominant to that of substructures (see below). The 
main features of the structure of $\Xi(\Delta x,\Delta v)$ coming from the smooth dark matter distribution can actually be explained by using 
the analytical result found for the isothermal
sphere (Eq.~\ref{MB_A}): 
\begin{itemize}
\item Small separations in phase space in Fig.~\ref{Fig_host} are denser in phase space due to the scaling of $\Xi$ with $\Delta x$ and
$\Delta v$; the former is related to the density profile making $\Xi\propto1/\Delta x$ for the isothermal sphere, while for CDM haloes the
scaling varies due to the radial change of the logarithmic density slope. 
\item Since $\Xi(\Delta x = {\rm cte}, \Delta v \ll \sigma_{\rm disp}) \sim~{\rm cte}$, the contours
of constant phase space density are vertical at small $\Delta v$. Once the velocity gets closer to the
velocity dispersion $\sigma_{\rm disp}$, the number of particles starts to decrease rapidly. For the isothermal sphere, 
the value of $\sigma_{\rm disp}$ is independent of radius and this is why the contours of constant
\psad~bend at the same value of $\Delta v$. For the CDM case however, the velocity dispersion is a function of radius, which explains why the 
``bend'' occurs at smaller values of $\Delta v$ for smaller values of $\Delta x$, where the \psad~is dominated by the colder and dense central regions 
of the the halo centre. 
\end{itemize}

\subsection{Redshift evolution and role of substructures}\label{evolution}

\begin{figure}
\center{
\includegraphics[height=8.0cm,width=8.0cm]{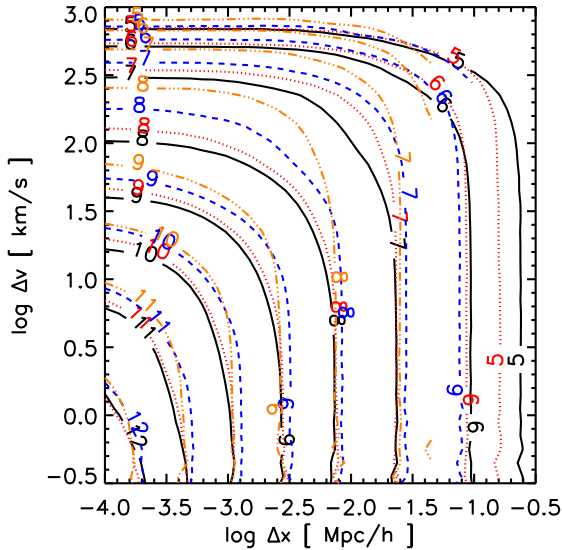}
}
\caption{Contours of the logarithm of the particle phase space average density (\psad) for the Aq-A-2 halo at different
redshifts: $z=0$ (solid black), $z=0.96$ (dotted red), $z=2.2$ (dashed blue) and $z=3.5$ (dashed-dotted orange). At each redshift, 
a sample from all particles within the main halo and its substructures is considered for the average.} 
\label{Fig_evol} 
\end{figure}

Fig.~\ref{Fig_evol} shows the redshift evolution of log(\psad) for the Aq-A-2 halo,
from $z=0$ (solid black) to $z=3.5$ (dashed-dotted orange). To construct this figure we computed \psad~with our code for a sample
of particles within the virialized region of the halo,
identified at the given redshift. The largest evolution 
with redshift occurs at intermediate to large separations in velocity while there is almost no evolution at small 
$\Delta v$. In order to have a better understanding of Fig.~\ref{Fig_evol}, we can model the Aq-A-2 halo as 
a superposition of a smooth main halo and a hierarchy of subhaloes down to the minimum reliable subhalo mass in the simulation: 
$\sim10^6$M$_{\rm \odot}$. At resolution level 2, a subhalo of this mass has $\gtrsim70$ particles, which is roughly the 
minimum number that we can trust in terms of global subhalo properties (see Figs. 26 and 27 of \citealt{Springel_08}). 

We already described the properties of the main halo in Section \ref{smooth}. For the subhaloes, we assume that they have scaling 
properties entirely defined by their mass $M_{\rm sub}$ \citep{Springel_08}:
$V_{\rm max}=10~{\rm km~s}^{-1} (M_{\rm sub}/3.37\times10^7~{\rm M}_{\rm \odot})^{0.29}$, 
$\delta_V/2=(V_{\rm max}/H_0r_{\rm max})^2=2.9\times10^4(M/10^8~{\rm M}_{\rm \odot})^{-0.18}$, where $\delta_V$ is a measure of concentration
and $r_{\rm max}$ is the radius where the circular velocity peaks; $H_0$ is the value of the Hubble constant today in the WMAP1 cosmology. We further assume 
that subhaloes also have Einasto density profiles
with a shape parameter that is mass-dependent \citep[see Fig. 1 of][]{Vera-Ciro_13}. Although the scatter in this relation is
large, we approximate: $\alpha_e(M_{\rm sub})={0.8,0.65,0.5,0.35}$ for $M_{\rm sub}={10^6,10^7,10^8,10^9}~{\rm M}_{\odot}$, respectively 
(and interpolating for masses in between). The other 
two parameters of the Einasto profile are then given by $r_{-2}=r_{\rm max}/a(\alpha_{e})$ and 
$\rho_{-2}=(V_{\rm max}/r_{-2})^2 / (b(\alpha_{e}) G)$, where $a(\alpha_e)$ and $b(\alpha_e)$ are functions of the shape parameter that are computed numerically  
by maximizing the circular velocity profile calculated from the enclosed mass:
\begin{eqnarray}\label{mass_einasto}
  M(r)&=&\frac{4\pi r_{-2}^3\rho_{-2}}{\alpha_e}{\rm exp}\left(\frac{3{\rm ln}\alpha_e+2-{\rm ln}(8)}{\alpha_e}\right)\times\nonumber\\
  &&\gamma\left(\frac{3}{\alpha_e},\frac{2}{\alpha_e}\left(\frac{r}{r_{-2}}\right)^{\alpha_e}\right)
\end{eqnarray}
where $\gamma(c,x)$ is the lower incomplete gamma function. For example, for $M_{\rm sub}=10^6~{\rm M}_{\odot} (\alpha_e=0.8)$, $a=1.756$ and $b=9.35$. 
Each subhalo has a cutoff radius $r_{\rm cut}$ which, by consistency, is given by the radius where the enclosed mass is equal to $M_{\rm sub}$. 
The velocity distribution in subhaloes is treated in the same way as in the main host with the proper scaling given by $\sigma_{\rm max}(M_{\rm sub})$ 
(see Sec. \ref{smooth}, below Eq. \ref{einasto}).

Finally, we assume that \psad~is mainly given by correlations among particle pairs within the main halo and within subhaloes,
i.e., we only consider the 1-(sub)halo term as is typically known in the halo model. The 1-subhalo term  
in a given mass range is weighted by the subhalo mass fraction relative to the host, which we obtain using the following fit to the
subhalo mass function \citep{BK_10,Gao_11}:
\begin{eqnarray}\label{sub_mf}
  \frac{dN}{d{\rm ln}\mu}=&&\nonumber\\
  \left(\frac{\mu}{\tilde{\mu_1}(z)}\right)^{a^\ast}{\rm exp}\left[\left(\frac{-\mu}{\mu_{\rm cut}(z)}\right)^{b^\ast}\right]
  \left(b^{\ast}\left(\frac{-\mu}{\mu_{\rm cut}(z)}\right)^{b^\ast} - a^{\ast}\right)
\end{eqnarray}
where $\mu=M_{\rm sub}/M_{200}$, and $a^\ast=-0.94$ and $b^\ast=1.2$ fit the simulation data very well (both the Aquarius haloes, 
and those from a larger box simulation, 
the Millennium II, \citealt{BK_09}) at all redshifts for a large range of host halo masses with slowly
varying values of $\tilde{\mu_1}(z)$ and $\mu_{\rm cut}(z)$ with redshift \citep[see Eq. 3 of][]{Gao_11}. 

\begin{figure}
\center{
\includegraphics[height=8.0cm,width=8.0cm]{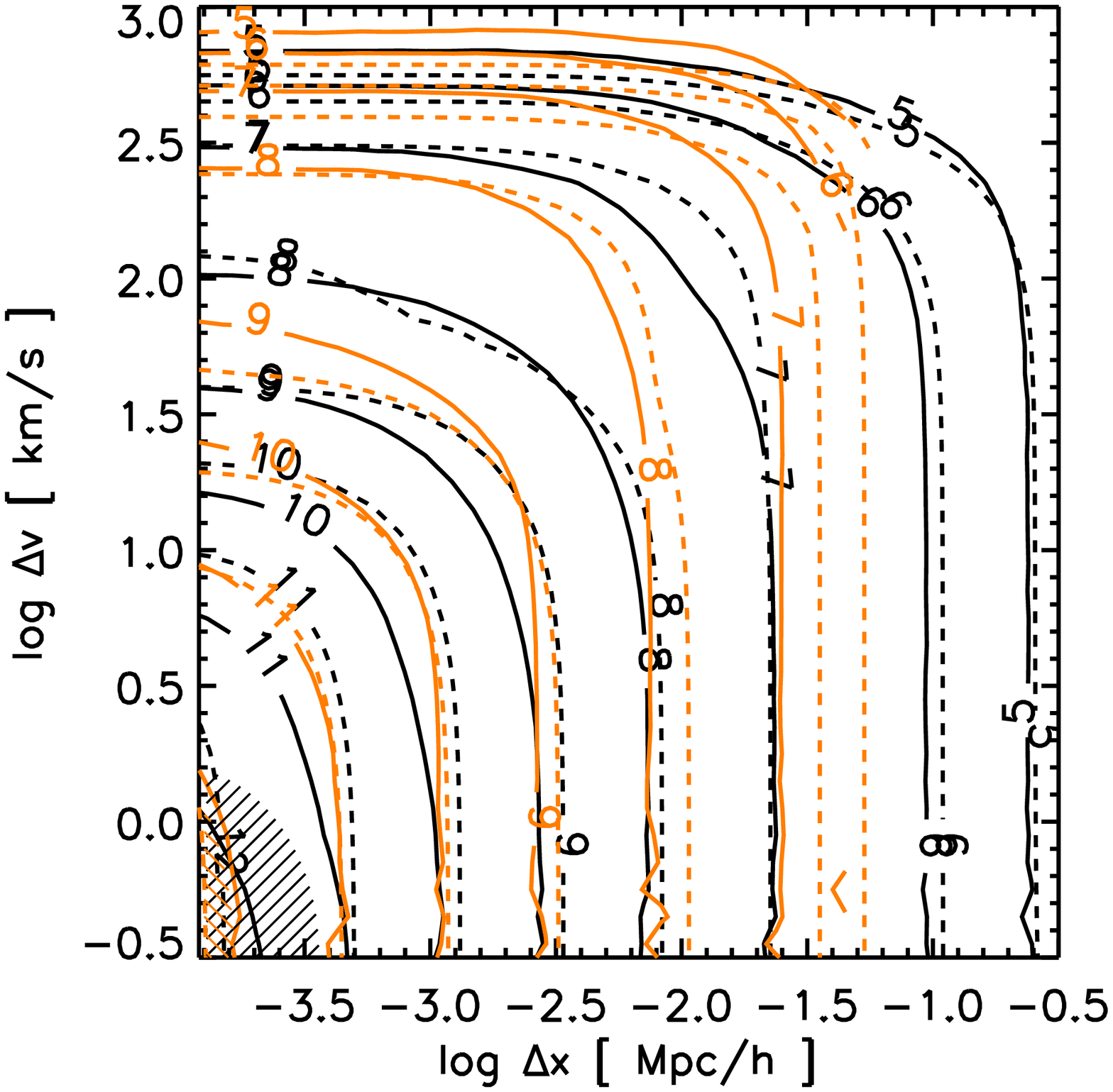}
}
\caption{Contours of the logarithm of the particle phase space average density (\psad) for the Aq-A-2 halo (solid) and for the halo model
described in the text (dashed) at two different redshifts: $z=0$ (black) and $z=3.5$ (orange). The shaded areas near the left corner 
encompass the region where the relative convergence between levels 3 and 2 is worse than $\sim1.5$ (see Appendix \ref{conv}).} 
\label{Fig_comp} 
\end{figure}

\begin{figure*}
\begin{tabular}{|@{}l@{}|@{}l@{}|}
\includegraphics[height=8.0cm,width=8.0cm]{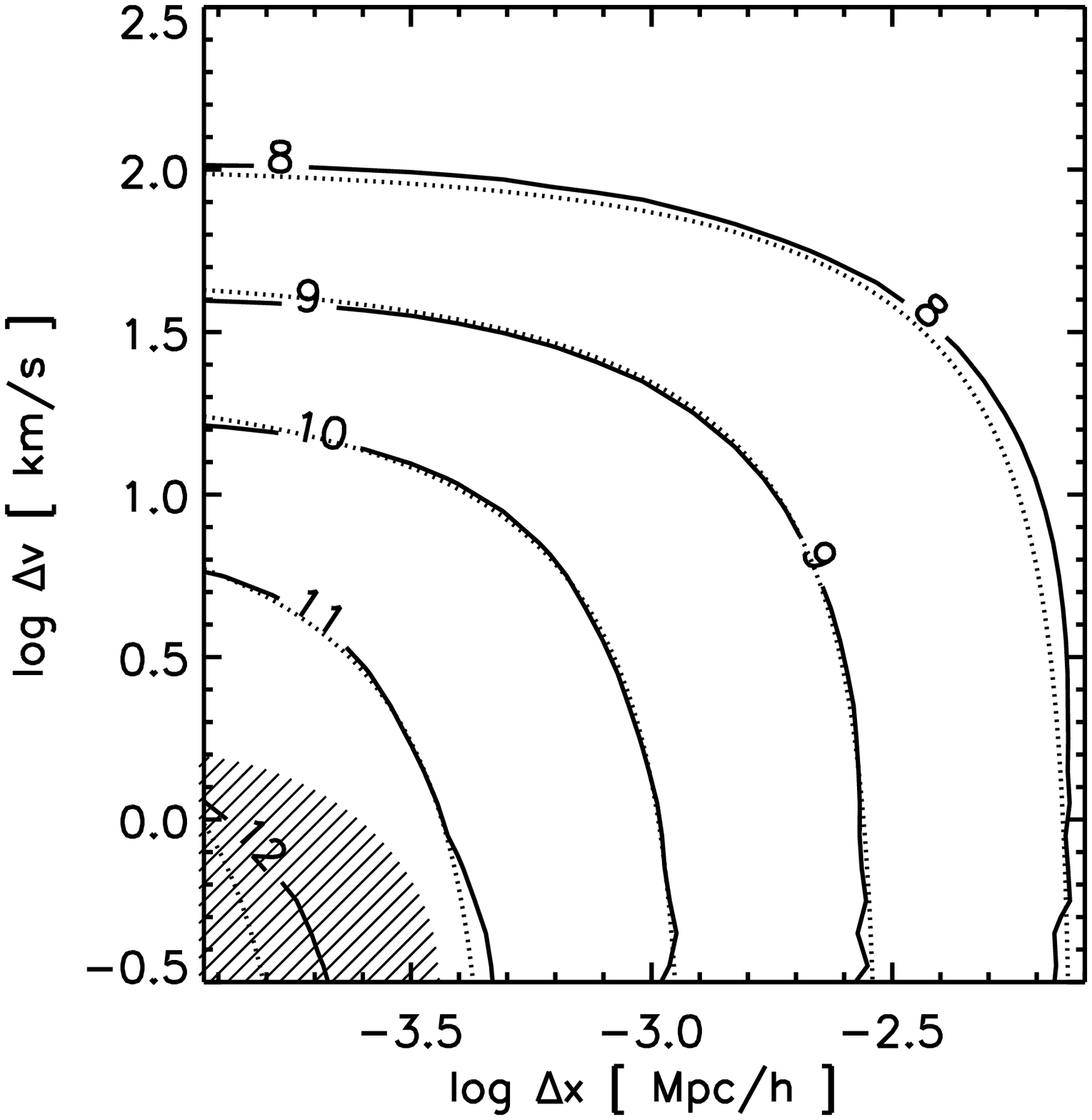} &
\includegraphics[height=8.0cm,width=8.0cm]{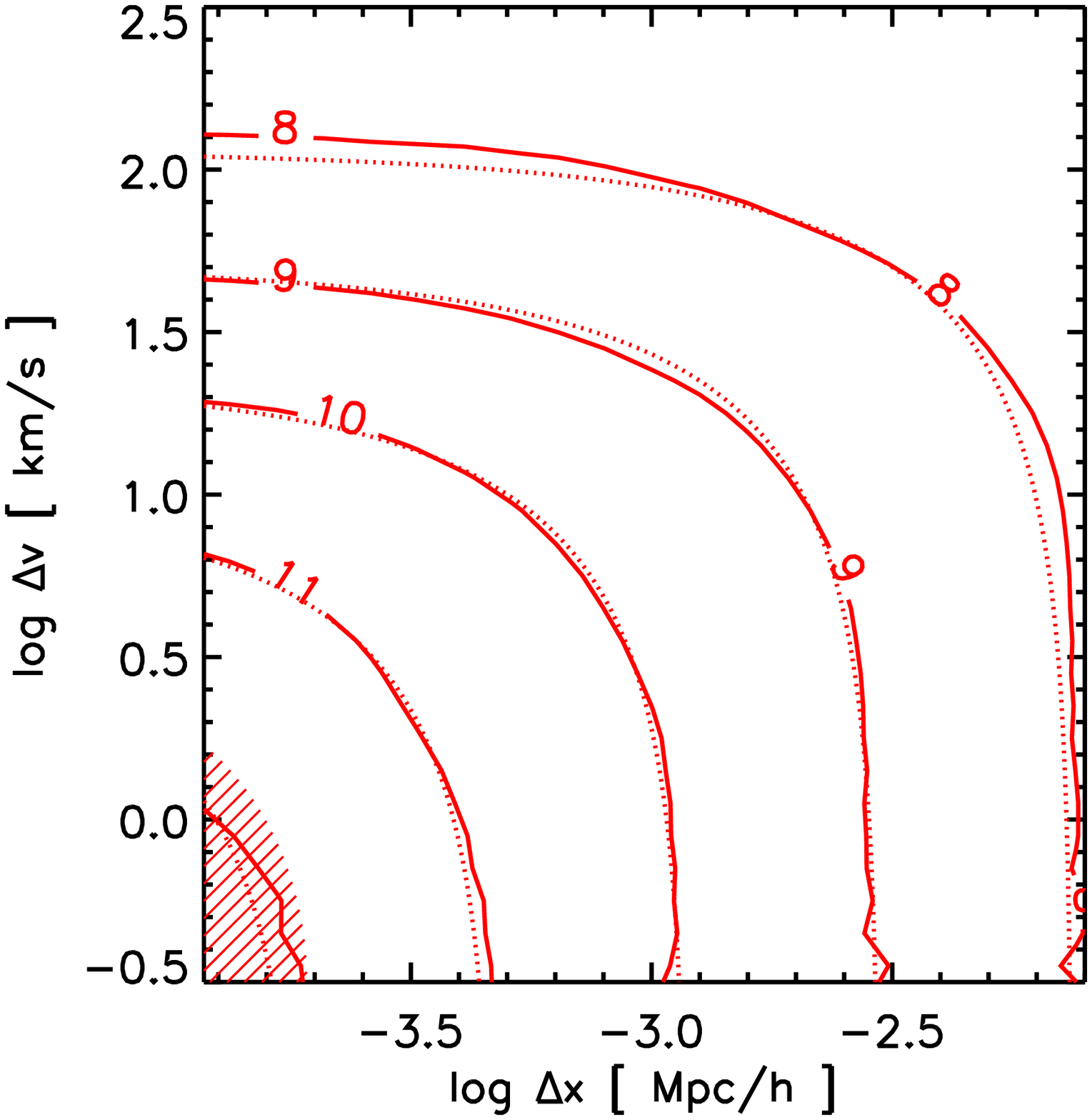} \\
\includegraphics[height=8.0cm,width=8.0cm]{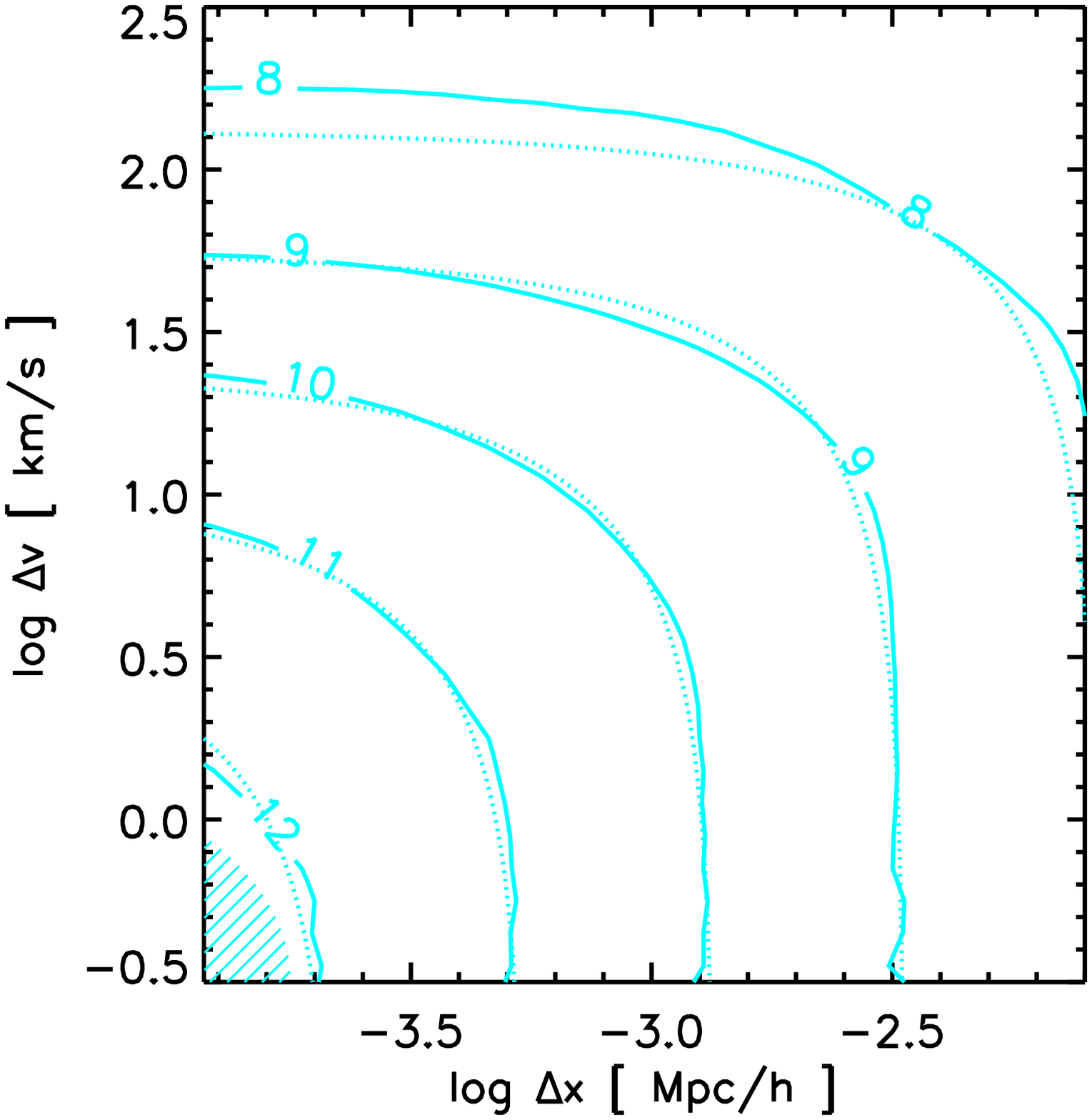} &
\includegraphics[height=8.0cm,width=8.0cm]{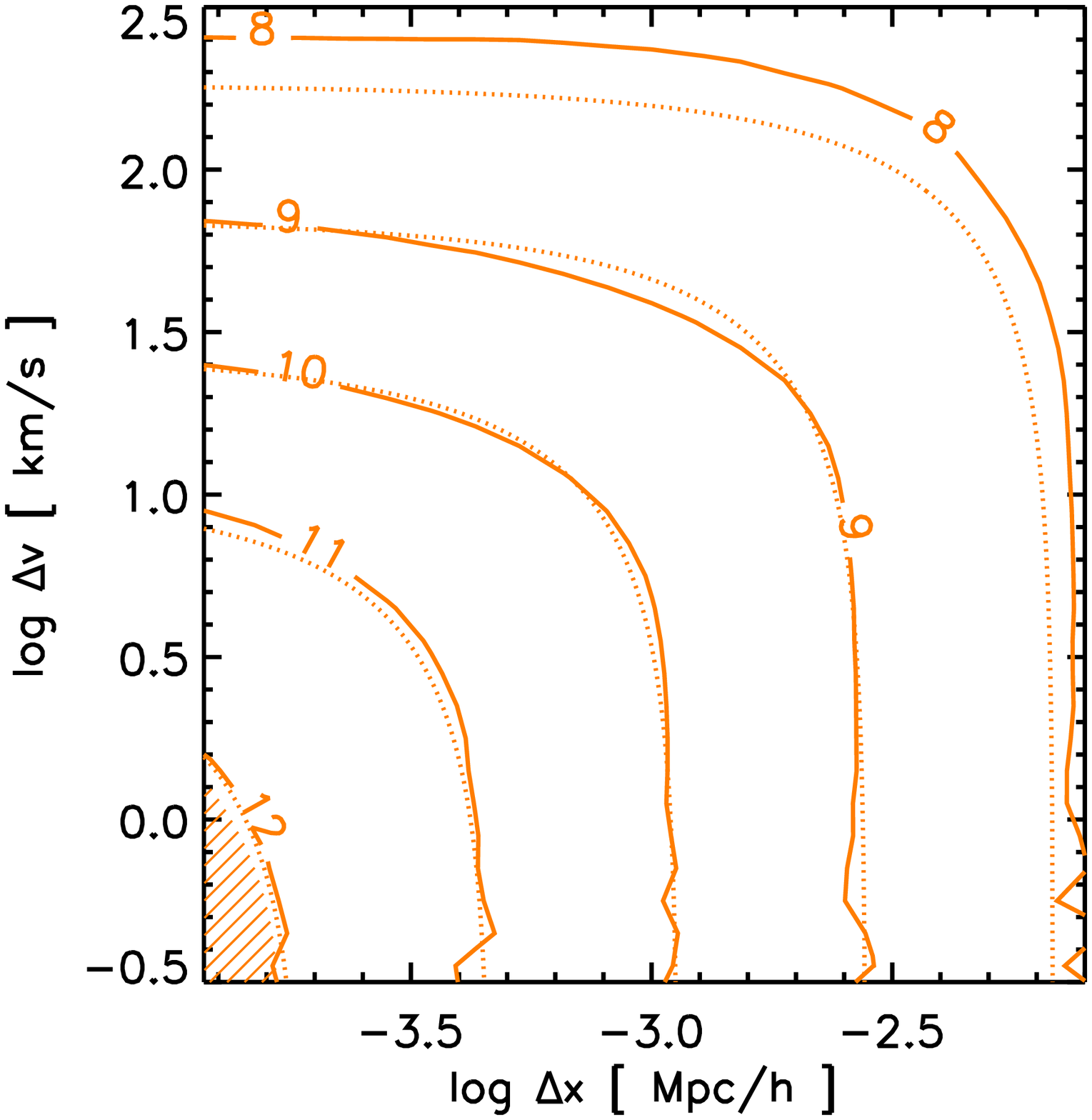} \\
\end{tabular}
\caption{Contours of the logarithm of the particle phase space average density (\psad) of the Aq-A-2 halo (solid) and for the fitting function
defined in Eqs.~\ref{lame}-\ref{lame_2} (dashed) with the parameters given in Table \ref{table_fit}. The different panels show
different redshifts, clockwise from top left: $z=0$ (black), $z=0.95$ (red), $z=2.2$ (cyan) and $z=3.5$ (orange).} 
\label{Fig_comp_fit} 
\end{figure*}

The particle phase space average density from the main host and its subhaloes is then given by:
\begin{eqnarray}
  \Xi(\Delta x, \Delta v) = \Xi_{\rm host}(\Delta x, \Delta v; M_{\rm host}) + \nonumber\\ 
  \int_{M_{\rm min}}^{M_{\rm max}}dM_{\rm sub}\frac{dN}{dM_{\rm sub}}\left(\frac{M_{\rm sub}}{M_{200}}\right)\Xi_{\rm sub}(\Delta x, \Delta v ; M_{\rm sub})
\end{eqnarray}
where $M_{\rm  min}$ is the minimum subhalo mass that is resolved in Aq-A2 ($\sim10^6{\rm M}_{\rm \odot}$)
and $M_{\rm max}$ is the largest subhalo in the simulation\footnote{We take $M_{\rm max}=5\times10^9{\rm M}_{\rm \odot}$. There are a 
few subhaloes above this mass, but they make a minimal contribution.}. We compute $\Xi_{\rm host}$ and $\Xi_{\rm sub}$ by
solving numerically Eq.~(\ref{2pcf_eq}) (due to the symmetries, the integral is 4D at the end: one radial and one angular in both, space and
velocity; the latter are fully analytical if we assume no cutoff to the velocity distribution). Fig.~\ref{Fig_comp} shows the comparison between the subhalo model described in this 
Section (dashed) and the simulation results (solid) at $z=0$ (black) and $z=3.5$ (orange). To model the smooth component of the latter, 
we assume that the main halo at this time is well 
represented by the same density profile as in $z=0$ but with $r_{200}(z = 3.5) \simeq 40$ kpc (with an enclosed mass  $\sim0.25M_{200}(z=0)$), i.e., we assume that the central region of the halo at $z=0$ was already in place 
at $z=3.5$, and the outer  boundary is simply set by $\rho_{\rm crit}(z)$. 

The model gives a good description of the simulated data at large and intermediate separations, with the difference mainly
caused by the use of a Maxwellian velocity distribution instead of a self-consistent one corresponding to the
Einasto profile. The agreement at
small separations is poorer, likely due to a combined effect of the missing ingredients in the subhalo model, such
as sub-substructures and a radial dependent concentration-mass relation. Nevertheless, the model is accurate to within a factor of $<2-3$ in most of the phase space region 
where convergence is good (outside the shaded areas in Fig.~\ref{Fig_comp}; see Appendix \ref{conv}). 

Notice that most of the evolution of \psad~with redshift occurs at large separations in velocity where the smooth halo component is dominant. 
This dominance happens roughly at $\Delta v\sim100$~km/s (see Fig.~\ref{Fig_host}). The model we have described in this section roughly accounts for
this evolution at small $\Delta x$: the dense centre of the halo is already in place at $z=3.5$ and further accretion of material only dilutes the 
coarse-grained average phase space density at lower redshifts since this new material is hotter than the cold central regions. 
Thus, the ``inside-out'' growth of the smooth component essentially accounts for most of the redshift evolution seen in Fig.~\ref{Fig_evol}, 
and is also related to the fact that the contours of constant phase space density expand in velocity but not in space at higher redshift.

\subsection{Fitting functions for \psad~on small scales}\label{fitting}

\begin{table}
\begin{center}
\begin{tabular}{ccccc}
\hline
Redshift  & $q_X({\rm Mpc}/h)$ & $\alpha_X$ & $q_V({\rm km/s})$ & $\alpha_V$ \\
\hline
\hline
0.0    &   $11.82$  & $-0.4$ & $4.5\times10^4$    & $-0.33$                                  \\  \hline
0.95   &   $11.82$  & $-0.4$ & $7.2\times10^4$    & $-0.35$                                  \\  \hline
2.2    &   $12.0$  & $-0.395$ & $1.2\times10^5$    & $-0.37$                                   \\  \hline
3.5    &   $9.0$  & $-0.415$ & $3.8\times10^5$    & $-0.415$                                   \\  \hline
\hline
\end{tabular}
\end{center}
\caption{Values of the fitting parameters in Eq.~\ref{lame_2}. Together with Eq.~\ref{lame}, these parameters provide a
good description of the \psad~at small $(\Delta x, \Delta v)$ (see Fig.~\ref{Fig_comp_fit}).}
\label{table_fit} 
\end{table}

The model described in the previous section provides a good description of \psad~on large scales, but it is less accurate
on small scales, where subhaloes dominate. We have found that, in this regime, a more precise description of the shape of 
\psad~is given by superellipses (Lam\'e curves). This functional shape is motivated by the stable clustering hypothesis and is the subject
of an extended study in paper II of this series \citep{Zavala_13}:
\begin{equation}\label{lame}
\left(\frac{\Delta x}{\mathcal{X}(\Xi)}\right)^{\beta} + \left(\frac{\Delta v}{\mathcal{V}(\Xi)}\right)^{\beta} = 1,
\end{equation}
where $\mathcal{X}(\Xi)$ and $\mathcal{V}(\Xi)$ are the generalised axes of the superellipse. A reasonable fit to the simulated
data at all redshifts (Fig. \ref{Fig_comp_fit}) is accomplished by the following parametrisation of the fitting parameters:
\begin{eqnarray}\label{lame_2}
\beta=\beta_0+\beta_1(1+z)\nonumber\\ 
\mathcal{X}(\Xi)=q_X(z)\Xi^{\alpha_X(z)}\\
\mathcal{V}(\Xi)=q_V(z)\Xi^{\alpha_V(z)}\nonumber
\end{eqnarray}
where $\beta_0=0.67$ and $\beta_1=0.08$, while the values of the rest of the parameters are given in Table 
\ref{table_fit}. Notice that with the exception of $q_V$ all parameters change by $\lesssim33\%$ up to $z=3.5$.
We should emphasize that the fitting function (Eq.~\ref{lame}) is primarily intended to describe \psad~ in the small separation regime, which 
is dominated by the halo substructure, as opposed to the large separation regime, which is dominated by the smooth halo component 
(see Fig. \ref{Fig_host}).

\begin{figure}
\center{
\includegraphics[height=8.0cm,width=8.0cm]{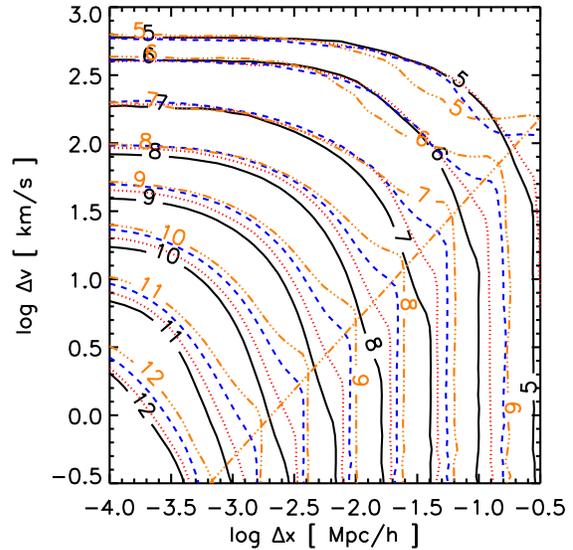}
}
\caption{Contours of the logarithm of the particle phase space average density (\psad) for
the Aq-A-2 halo at different redshifts: $z = 0$ (black), $z = 0.96$ (red), $z = 2.2$ (blue) and $z = 3.5$ (orange). The Lagrangian region
containing all the high-resolution particles at $z = 0$ is considered for all redshifts. This region covers a larger volume (highly asymmetric) 
than previous plots: the largest sphere within this volume has a radius of $\sim3.3(0.7)~{\rm Mpc}$ at $z=0(3.5)$ in
physical coordinates from the center of the main halo. The mean density within this sphere is of $\mathcal{O}(\rho_{\rm crit})$. The
straight orange line correspond to the Hubble flow at $z=3.5$, which is clearly responsible from the ``bump'' feature appearing more prominently
at higher redshifts.}
\label{Fig_full} 
\end{figure}

\subsection{Environmental dependence }\label{vol_sec}

So far, we have only analysed the particle phase space average density (\psad) averaged within a specific volume: the one defined by the virial 
radius of the halo. By changing
the volume where this average is made, we expect to observe changes in the structure of \psad. First, we show in Fig.~\ref{Fig_full} the 
redshift evolution of \psad~averaged over a larger volume, that of the Lagrangian region encompassed by all high-resolution particles at $z=0$.
This region is non-spherical, but the volume it covers is mostly contained within a sphere of roughly $3.3(0.7)~{\rm Mpc}$ at $z=0$ ($z=3.5$)
in physical coordinates from the centre of the main halo. This is well beyond the virial radius, and thus, the average of \psad~is a combination
of the contribution of the main halo (including its substructures) and a plethora of low mass field haloes (with their own substructure) and unclustered matter 
in the surroundings of the main host. 
Notice that, albeit the evolution of \psad~in this case is stronger than in the virialized region (Fig.\ref{Fig_evol}),
there is still only little evolution of \psad~at small separations.

\begin{figure*}
\begin{tabular}{|@{}l@{}|@{}l@{}|}
\includegraphics[height=8.0cm,width=8.0cm]{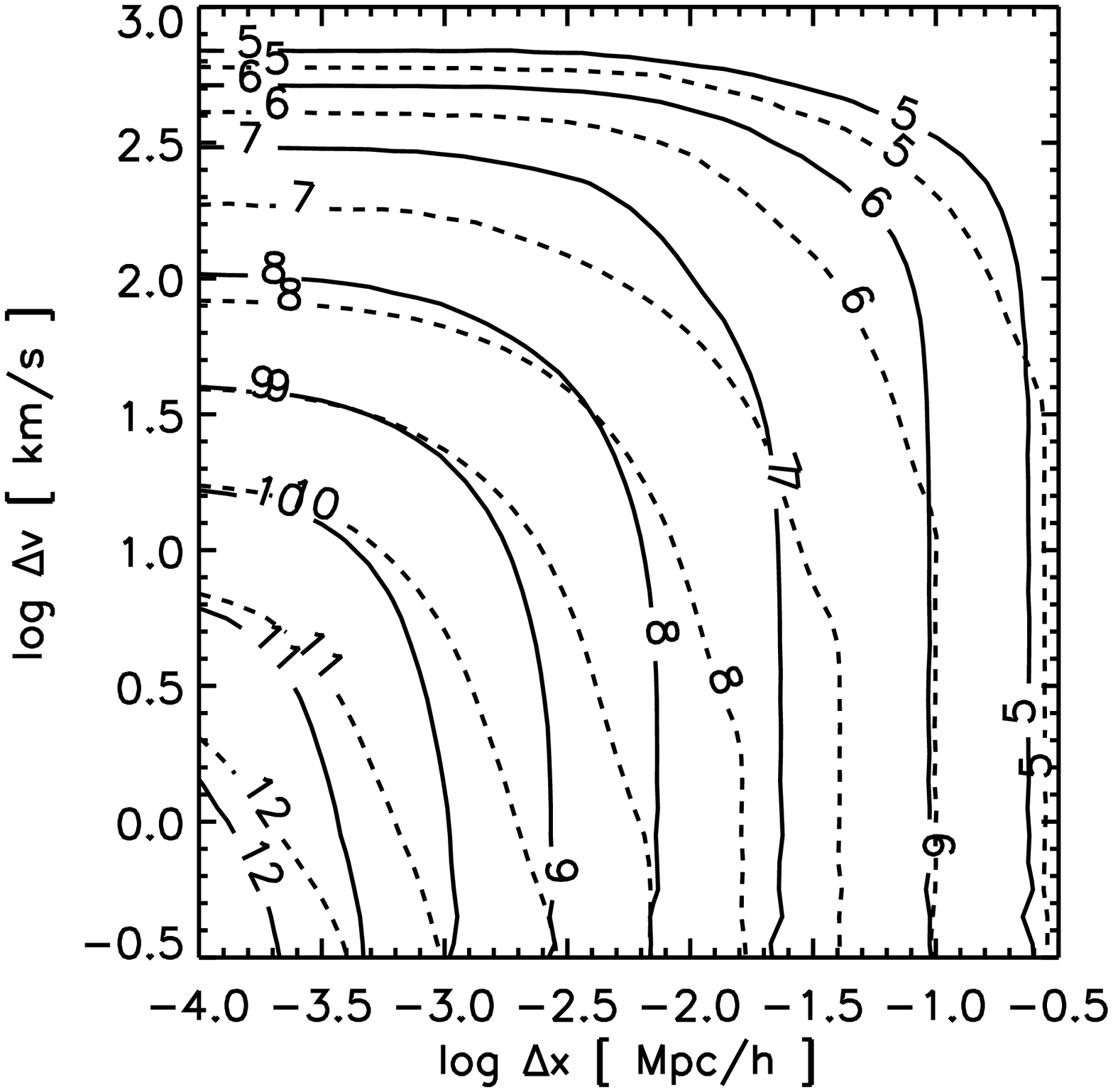} &
\includegraphics[height=8.0cm,width=8.0cm]{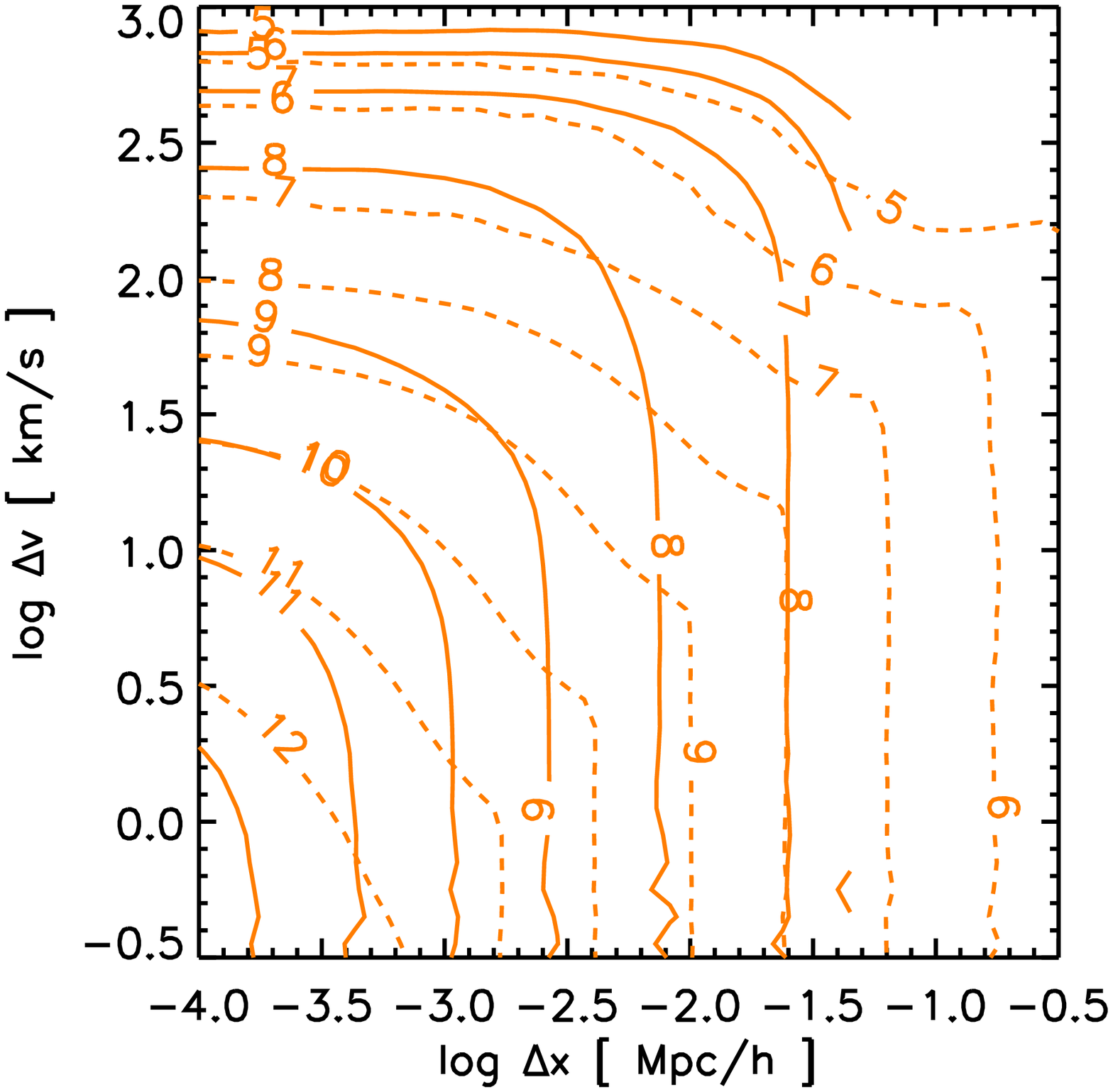} \\
\end{tabular}
\caption{Contours of the logarithm of the particle phase space average density (\psad) for
the Aq-A-2 halo averaged over different volumes: only particles within the main halo and its subhaloes (solid line), 
and over a larger volume defined by all the high resolution particles (see caption of Fig.~\ref{Fig_full} and text for details). 
The left panel is for $z=0$ and the right panel for $z=3.5$.} 
\label{Fig_comp_full_main_subs} 
\end{figure*}

\begin{figure*}
\begin{tabular}{|@{}l@{}|@{}l@{}|}
\includegraphics[height=8.0cm,width=8.0cm]{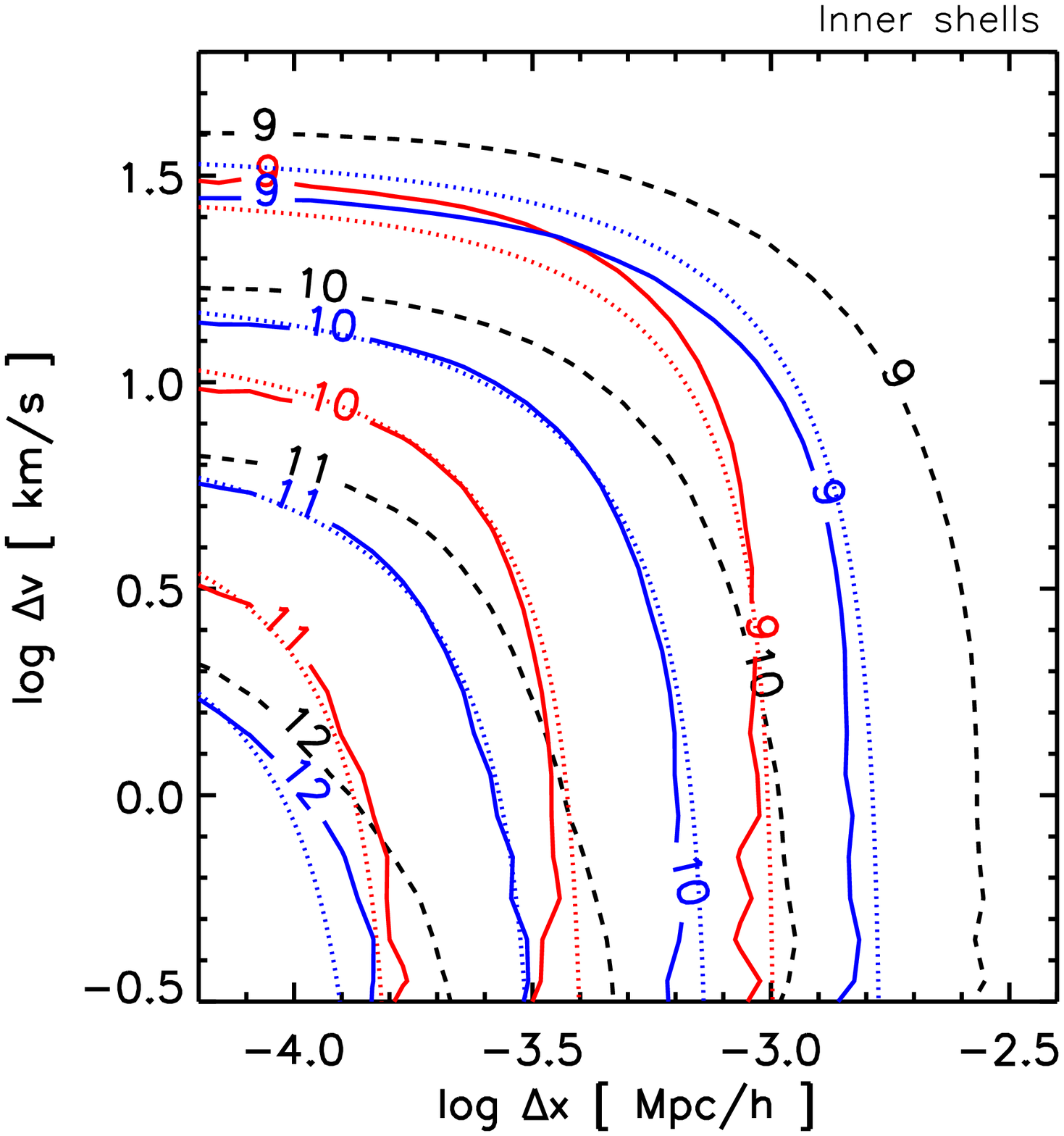} &
\includegraphics[height=8.0cm,width=8.0cm]{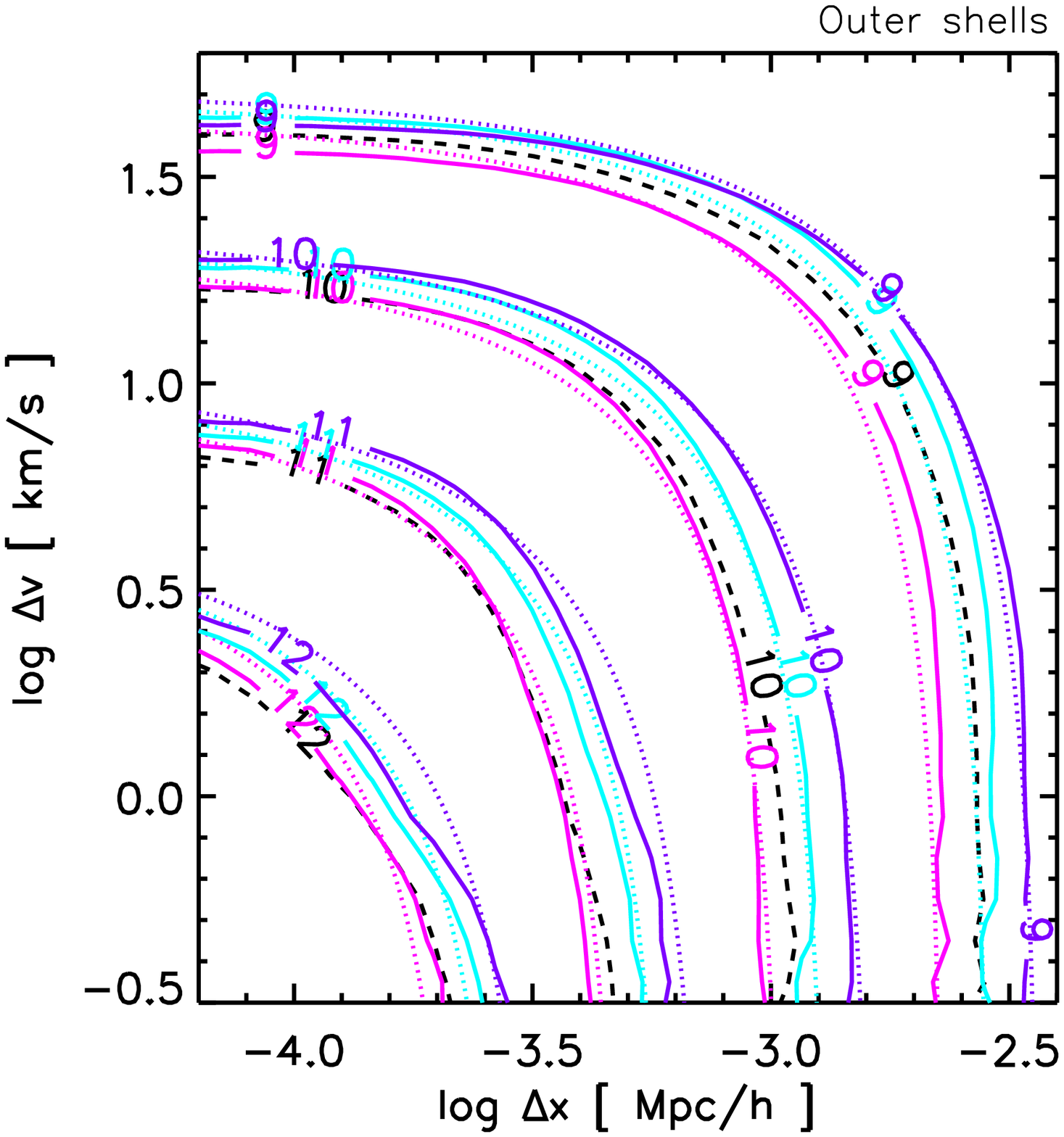} \\
\end{tabular}
\caption{Contours of the logarithm of the particle phase space average density (\psad) for the Aq-A-2 halo 
at $z=0$ for samples of particles taken at different radial shells from the halo centre. {\it Left panel}:
$0<r/r_{200}<0.2$ (red) and $0.2<r/r_{200}<0.4$ (blue). {\it Right panel}: $0.4<r/r_{200}<0.6$ (magenta), $0.6<r/r_{200}<0.8$ 
(cyan), and $0.8<r/r_{200}<1.0$ (violet). A sample of eight million particles was used for each shell.
The dotted lines are fits given by Eqs.~\ref{lame}-\ref{lame_2} with the parameters shown in table 
\ref{table_fit_radial}. For reference, the dashed line is \psad~averaged over the whole halo (i.e., the black solid line in Fig.~\ref{Fig_evol}).} 
\label{Fig_radial_dep} 
\end{figure*}

The difference between the different volumes taken for the average can be clearly appreciated in Fig.~\ref{Fig_comp_full_main_subs} where
we show the contours of log(\psad) for $z=0$ (left) and $z=3.5$ (right). There are
important differences relative to the case we explored before: 
\begin{itemize}
\item Looking at Fig.~\ref{Fig_full}, we see that at large $\Delta v$ there is almost no redshift evolution possibly because
\psad~is dominated by particles outside the main halo that undergo little merging and/or tidal disruption. We should note that by taking the average of
\psad~over all velocity separations $\Delta v$, one is probing the spatial 2PCF. By doing so we also find a (near) lack of evolution, which might 
be taken as evidence for the original (position-space) stable clustering hypothesis \citep{Davis_77}. 
\item There is a clear ``bump'', more prominent at higher redshifts, where the contours of constant phase space density bend, related to 
the Hubble flow, which, at fixed physical separation, is more important at high redshifts. The latter is clearly demonstrated by
the straight orange line in Fig.~\ref{Fig_full} that shows the Hubble flow for $z=3.5$.
\item Interestingly, at small $\Delta v$, a less dense environment, increases the average value of \psad, presumably because the 
low-mass field haloes (which are substantially denser in phase space) contribute most to the average. 
That is, the 1-(sub)halo term in \psad~receives additional contributions from low-mass field haloes and their subhaloes, and also the 2-halo
term appears as a novel contribution, particularly for large $\Delta x$.
This is a likely explanation for most of the difference between 
the dashed and solid lines in Fig.~\ref{Fig_comp_full_main_subs}, and also explains why only the velocity axis evolves for the halo region, while both 
axes evolve for the larger region. Notice how for the smallest values of $\Delta x$, the volume average seems to 
converge in both cases.
\end{itemize}

Let us now consider denser environments for the average of \psad~by taking concentric shells of increasing volume within the halo. In particular, we take
five radial shells with the same thickness: $0.2~r_{200}$, and consider only the case at $z=0$. The left (right) panel of Fig~\ref{Fig_radial_dep}
shows the contours of log(\psad) for the first (last) two (three) shells. For reference we show in both panels the
result of the average within the virialized region of the halo with a black dashed line (corresponding to the black solid line of Fig.~\ref{Fig_evol}). 
Overall, there is a trend of increasing phase space density towards the periphery of the halo.
This is clearly connected to the increasing total number of subhaloes as a function of radius, 
since it is well known that the mean/characteristic phase space density of the main (or spherically averaged) 
halo decreases in the outer parts \citep[e.g.][]{Taylor_01}.
Closer to the halo centre, tidal stripping has been more effective in destroying subhaloes diluting the
average value of \psad. 
Subhalo abundance and the radial-dependent tidal stripping also explain the clear change in the shape of the iso-\psad~contours, with
the ones corresponding to the outter shells being rounder due to the increasing number of subhaloes, while the inner shells have a shape
closer to the smooth halo component (see contour plot in Fig.~\ref{Fig_host}). 
Notice that within the third shell (i.e. $0.4<r/r_{200}<0.6$), \psad~is very close to the average within the whole halo.

The changes in shape and normalization of \psad~for the different radial shells can be captured accurately by the fitting functions provided in
Eqs.~(\ref{lame}-\ref{lame_2}) with the parameters given in Table \ref{table_fit_radial}.

\begin{table}
\begin{center}
\begin{tabular}{cccccc}
\hline
$r/r_{200}$   & $\beta$ & $q_X({\rm Mpc}/h)$ & $\alpha_X$ & $q_V({\rm km/s})$ & $\alpha_V$ \\
\hline
\hline
0.0-0.2    & $1.0$   &   $4.06$   & $-0.40$ & $4.0\times10^4$    & $-0.35$                                  \\  \hline
0.2-0.4    & $0.9$   &   $2.97$   & $-0.36$ & $3.34\times10^4$     & $-0.33$                                  \\  \hline
0.4-0.6    & $0.8$  &   $2.63$  & $-0.34$ & $4.1\times10^4$    & $-0.33$                                   \\  \hline
0.6-0.8    & $0.775$ &   $3.29$   & $-0.34$ & $5.6\times10^4$      & $-0.34$                                   \\  \hline
0.8-1.0    & $0.75$   &   $5.14$   & $-0.35$ & $5.9\times10^4$    & $-0.34$                                   \\  \hline
\hline
\end{tabular}
\end{center}
\caption{Values of the parameters in Eq.~\ref{lame_2} for the fit to \psad~averaged over different radial shells centered in the Aq-A-2 halo as indicated in the
first column (see Fig.\ref{Fig_radial_dep}).}
\label{table_fit_radial} 
\end{table}

\begin{figure*}
\begin{tabular}{|@{}l@{}|@{}l@{}|@{}l@{}|}
\includegraphics[height=6.0cm,width=6.0cm]{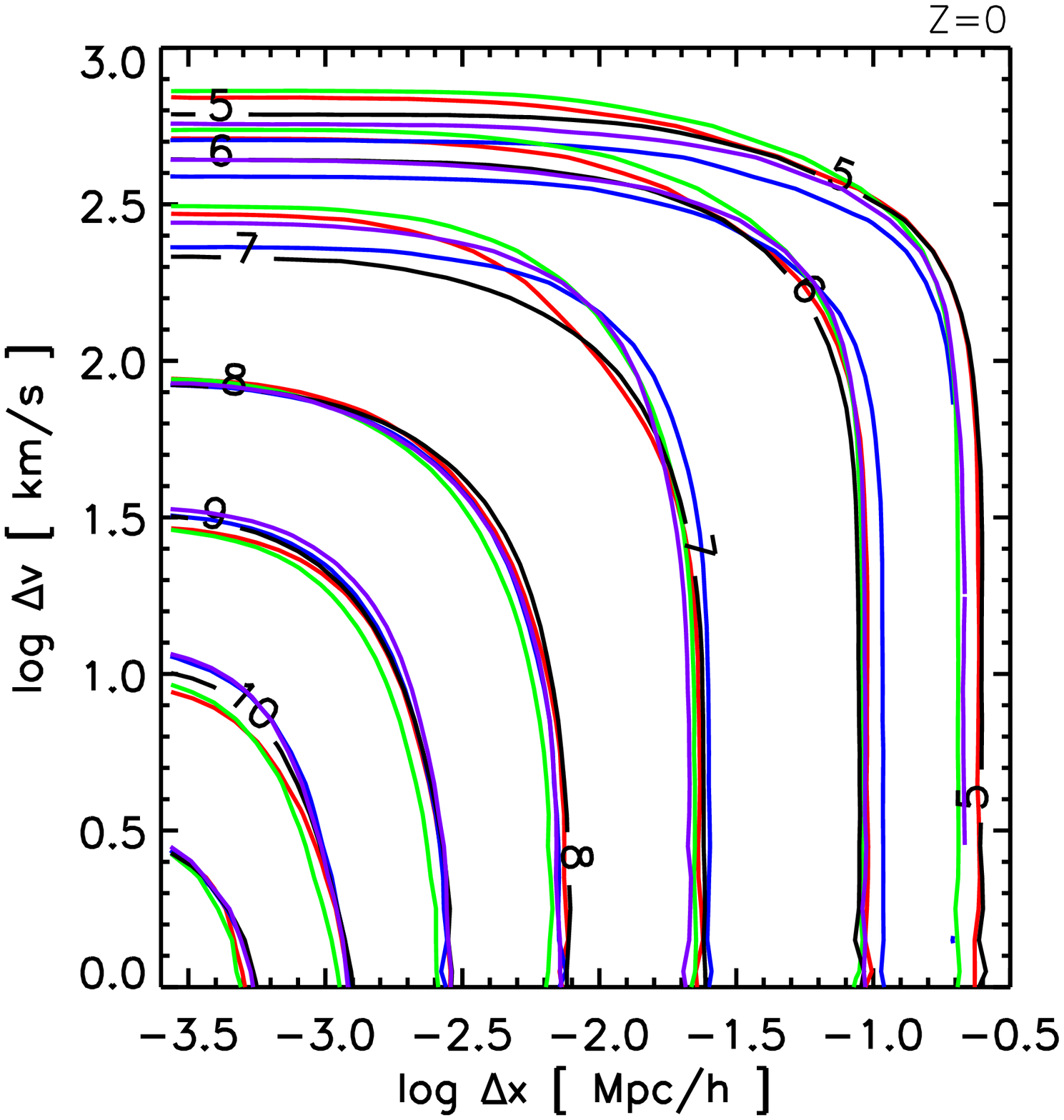} &
\includegraphics[height=6.0cm,width=6.0cm]{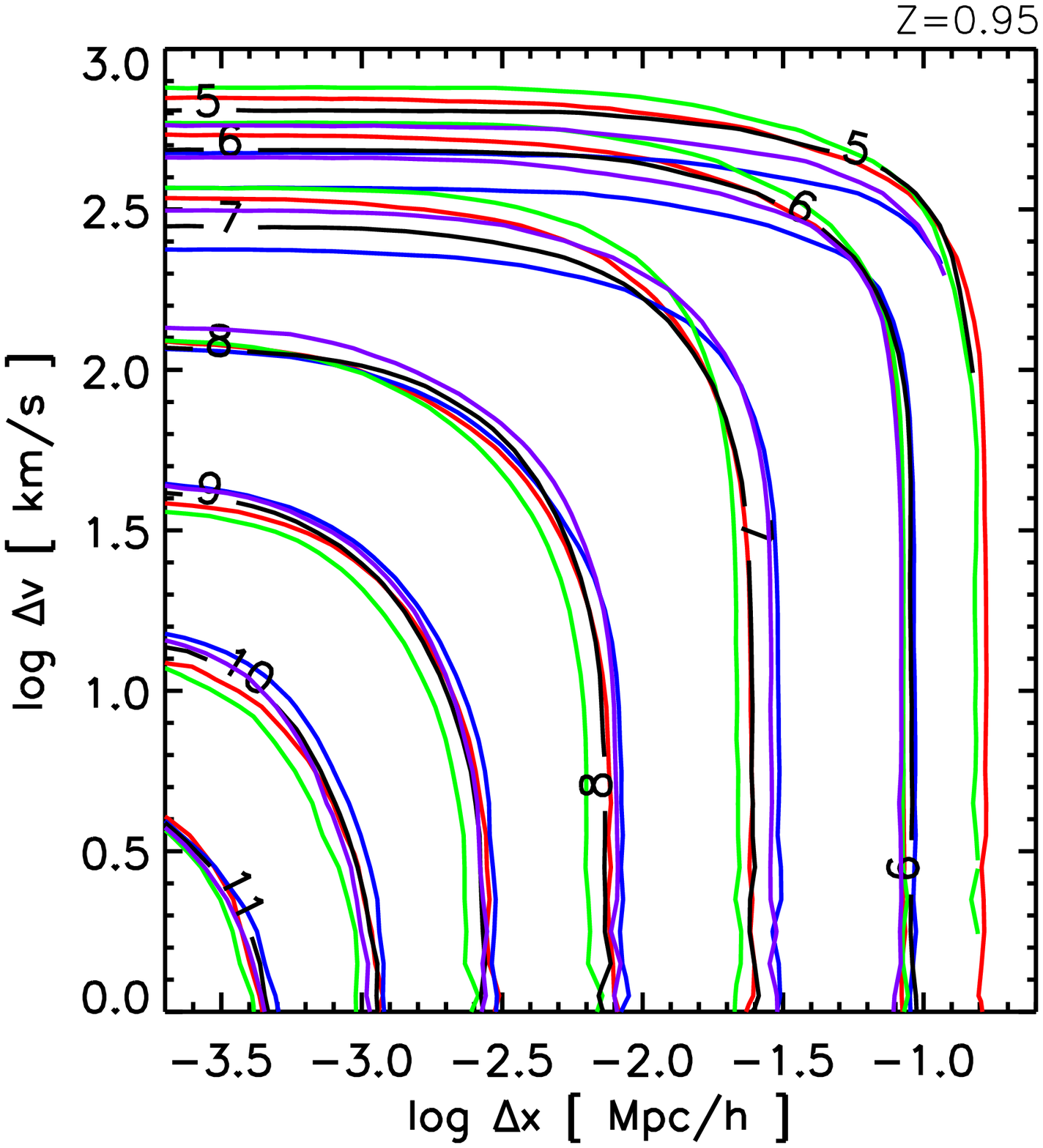} &
\includegraphics[height=6.0cm,width=6.0cm]{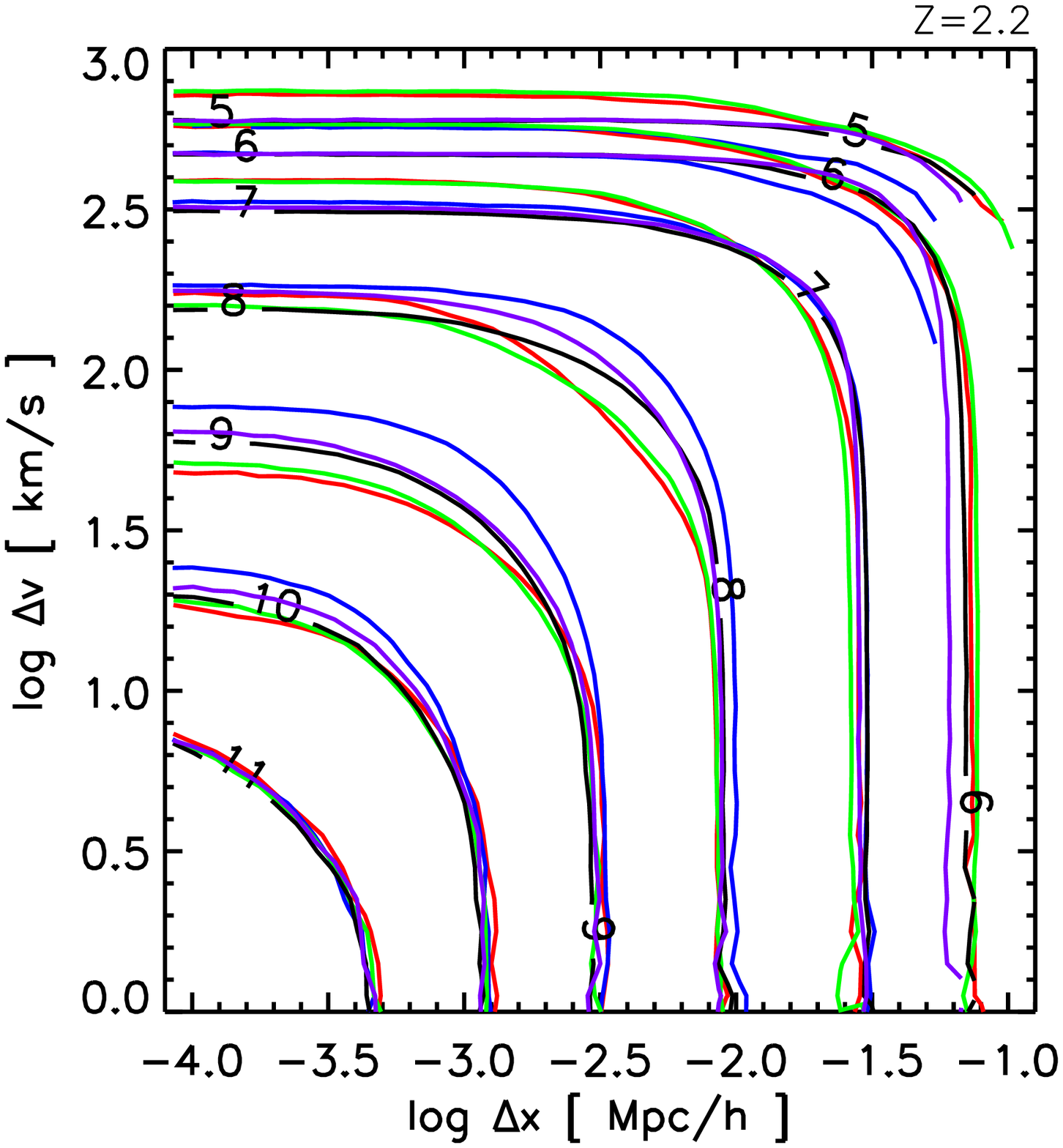} \\
\includegraphics[height=6.0cm,width=6.0cm]{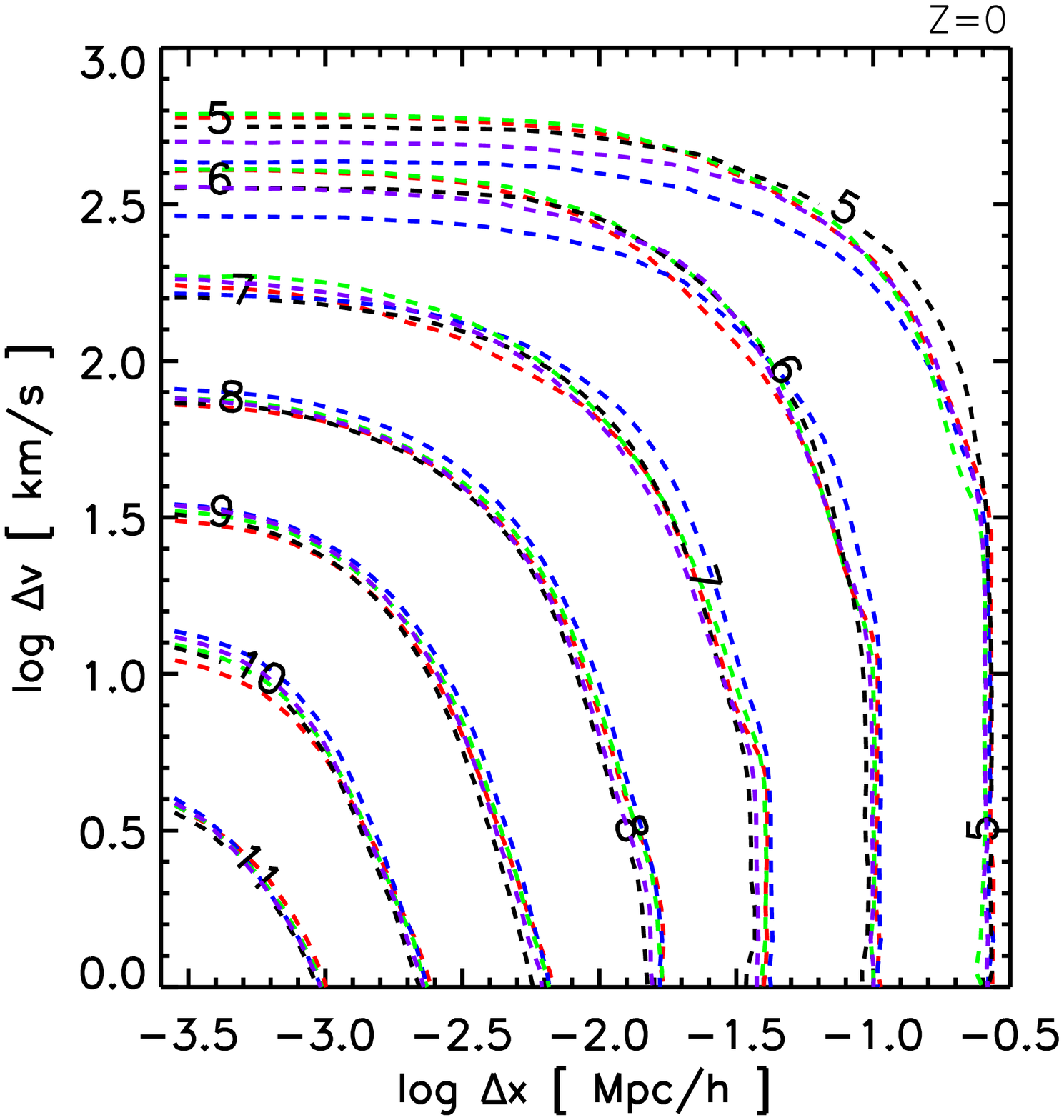} &
\includegraphics[height=6.0cm,width=6.0cm]{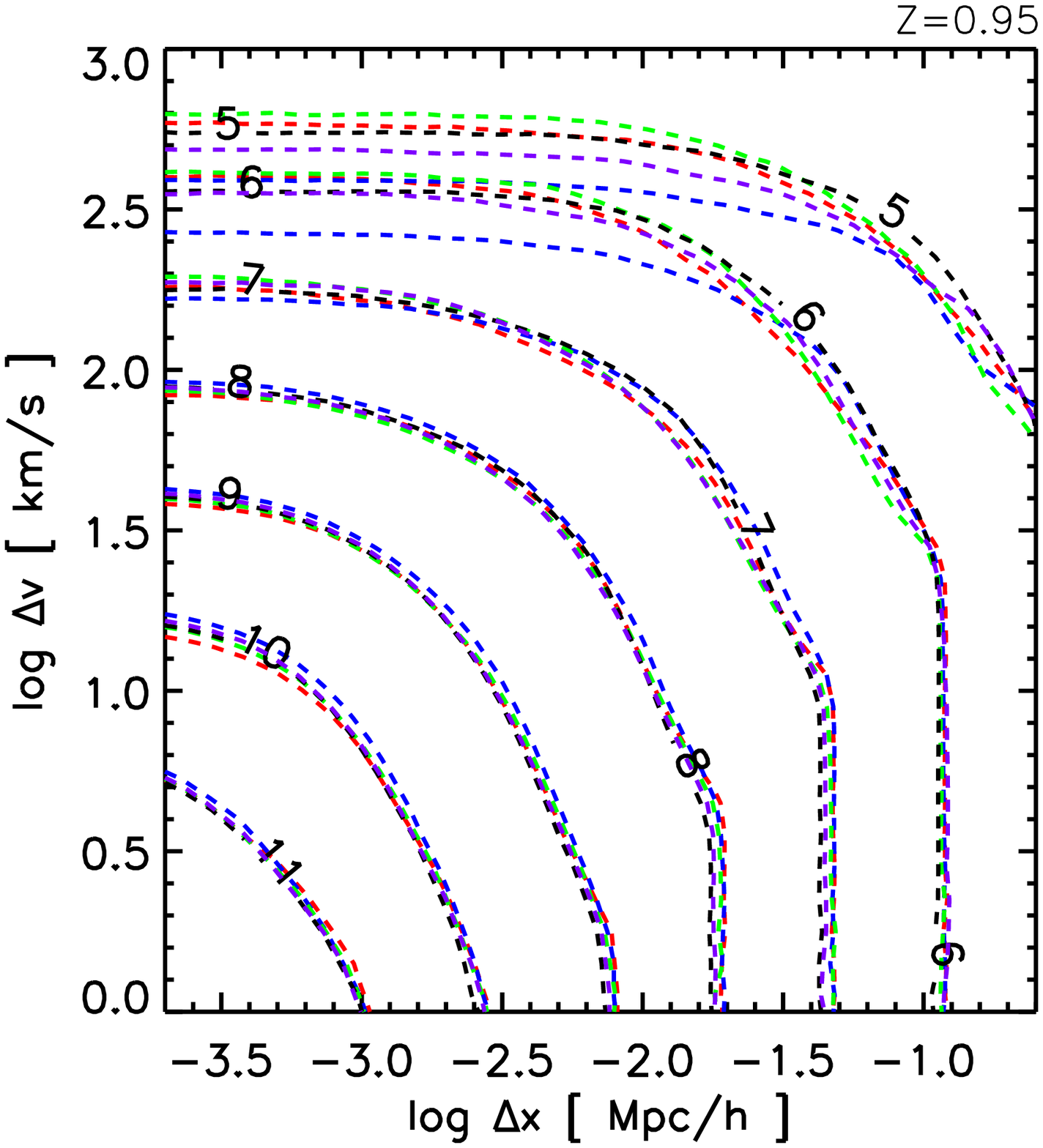} &
\includegraphics[height=6.0cm,width=6.0cm]{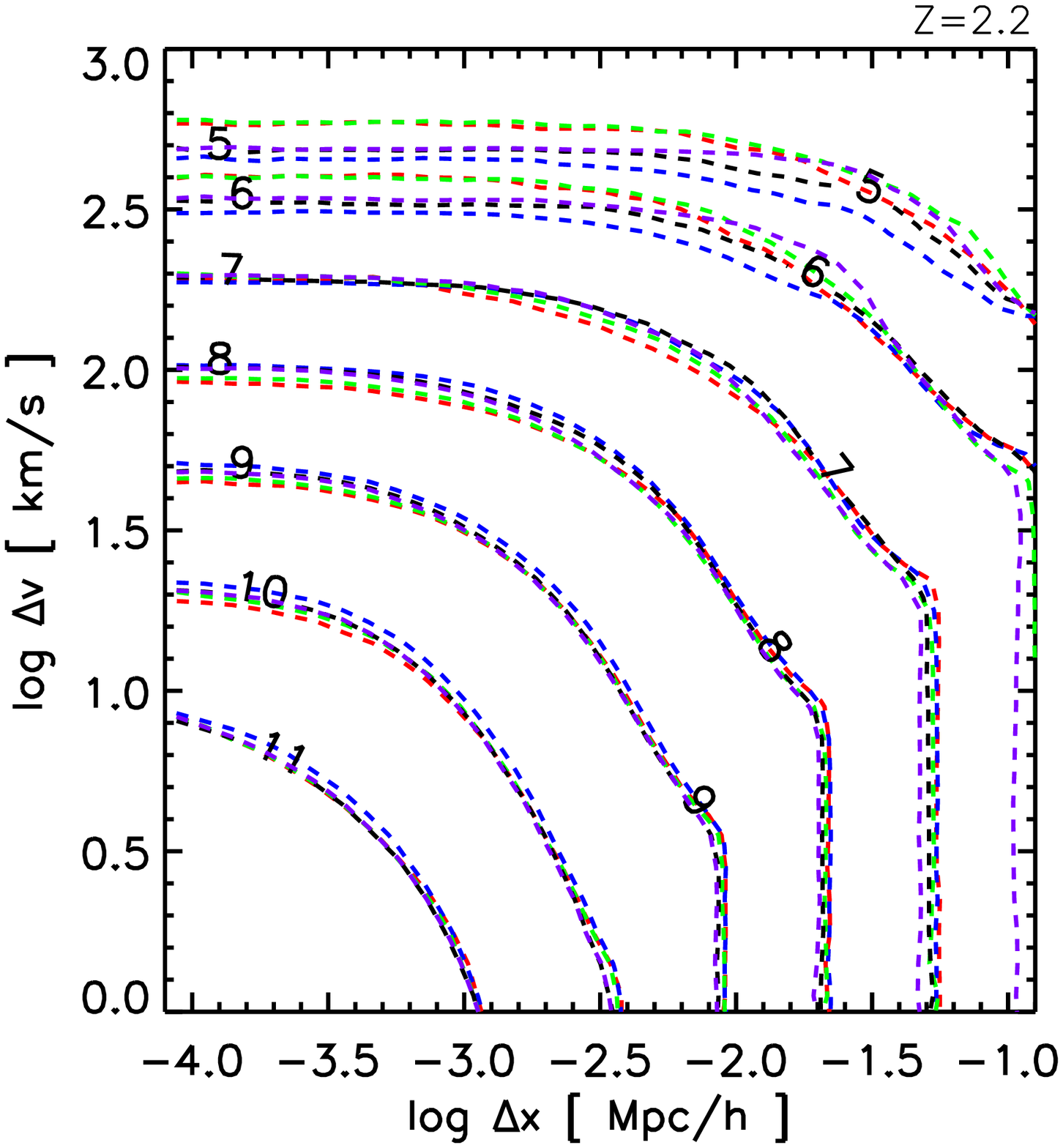} \\
\end{tabular}
\caption{Contours of the particle phase space average density (\psad) for the Aquarius haloes:
Aq-A (red), Aq-B (blue), Aq-C (green), Aq-D (black), Aq-E (violet)
at different redshifts: $z=0$ (left), $z=0.96$ (middle) and $z=2.2$ (right). 
In the upper panels, only particles within the main halo and its subhaloes are considered at each redshift while in the bottom panels, 
all the high-resolution particles at $z = 0$ are considered for all redshifts (see Fig.\ref{Fig_full} and Section \ref{vol_sec}).
In all cases the resolution level is 4. Note the change in the range of the horizontal axis in the different panels. } 
\label{Fig_diff} 
\end{figure*}

\subsection{Formation history: different Aquarius haloes}\label{mahs}

Fig.~\ref{Fig_diff} shows the variation of the particle phase space average density in the different Aquarius haloes at resolution level 4
at different redshifts according to the legends in the top right corner of each panel. 
These are MW-like haloes with different masses at $z=0$: $M_{200}\in(0.8,1.8)10^{12}~{\rm M}_\odot$ (see Table \ref{table_sims}), concentrations
that vary up to a factor of two, and very 
different mass aggregation histories representing a wide range of the full distribution expected for haloes of this mass in a cosmological volume 
\citep[see Figs. 2 and 3 of][for ease of comparison we have used their color scheme in Fig.~\ref{Fig_diff}]{BK_10}. The upper panel is for the volume average within the halo while the lower panel is for a larger volume given by
the Lagrangian region of the high-resolution particles at $z=0$. It is remarkable that despite the diverse merger histories, \psad~seems 
to be universal, particularly at small separations where the contours of constant log(\psad) of the different haloes are very close
to each other with small variations in redshift up to $z\sim2$. It is important to remark that at level 4 of resolution, 
\psad~is only converged for log(\psad)$\sim9$ (see Fig.~\ref{Fig_host_vs_all}), and thus the results at higher values of \psad, although suggestive,
should be taken with care. Nevertheless, this resolution is enough to probe the regime where subhaloes start to dominate and point towards the
universality we mentioned before.

Although we do not explore this further, at large $\Delta v$, the difference in the 
contours of phase space density at a given redshift is due to the different properties of the smooth distribution, while variations across different
redshifts seem to be correlated with the mass aggregation history of the halo. The fact that the difference between different haloes is smaller
for the larger volume (lower panels of Fig.~\ref{Fig_diff}) seems to support the latter. 

\section{Summary and Conclusions}\label{conclusions}

We provide a different perspective on the clustering of dark matter in phase space by introducing the Particle Phase Space
Average Density: \psad$\equiv\Xi(\Delta x, \Delta v)$. This quantity is defined as the 2D two-point correlation function (2PCF) in phase space 
divided by the mean phase space density within the phase space volume where the average is made (see Eqs.~\ref{2pcf_eq_0}-\ref{2pcf_eq}). It is thus
a generalization of the concept of the 2PCF in position space as a measure of the number of particle pairs separated by
a distance $\Delta x$ in space and $\Delta v$ in velocity space. 

We analyse the structure of \psad~in galactic haloes by using the simulation suite of the Aquarius project \citep{Springel_08} and
find the following main results:
\begin{itemize}
\item The structure of \psad~averaged within the virialized halo is divided into two distinct regions corresponding to the 
  smooth dark matter distribution at large separations in phase space and to ``all'' the hierarchy of substructures at small separations 
  (see Fig.~\ref{Fig_host}).
\item Most of the evolution of \psad~across time occurs at large separations and is roughly understood by the ``inside-out'' growth of the
  smooth dark matter halo. 
\item In the regime where gravitationally bound substructures dominate, there is a remarkable nearly universal behaviour of \psad~across: 
  \begin{itemize}
    \item {\it Time} (Fig.~\ref{Fig_evol}), with variations of a factor $\lesssim3$ between $z=0$ and $z=3.5$, while the average density within the halo 
      changes by a factor of $\sim30$.
    \item {\it Ambient density} (Figs.~\ref{Fig_comp_full_main_subs}-\ref{Fig_radial_dep}), varying by a factor of $\sim30$ at the most, while the average density changes by nearly four orders of magnitude. 
    \item {\it Mass accretion histories} (Fig.~\ref{Fig_diff}). The structure of \psad~changes only slightly for the five MW-size haloes
      we analyzed (typically much less than a factor of 2), despite of their diverse mass accretion histories, and subhalo 
      distributions.
  \end{itemize}
\item In the small scale regime we found that \psad~can be described by a family of superellipse contours in the $(\Delta x, \Delta v)$ plane with axes that scale with the phase space density as power laws (see Eqs.~\ref{lame}-\ref{lame_2}). By introducing slow variations in redshift
  and halo-centric distance for the parameters in the superellipse, we are able to obtain good fits to the simulation data in all cases. 
\end{itemize}

The functional shape that fits the structure of \psad~at small separations is inspired by a model based on  an extension into phase space of 
the stable clustering hypothesis, originally introduced by \citet{Davis_77} in position space. In a companion paper \citep{Zavala_13}, we describe this 
model in great detail and show how, coupled with a simplified tidal disruption scenario, it can describe the simulation data quite well.

Although we have shown that \psad~is sensitive to cold small scale gravitationally bound subhaloes, we want to remark 
that \psad~is not a substitute for a substructure finder, but rather an alternative way of studying dark matter 
clustering. This is similar to contrasting a measurement of the galaxy two-point correlation function, with identifying 
groups and clusters of galaxies, and their statistics. As is well known in the studies of large scale structure, the 
two contain complementary (but to some extent overlapping) information.

The structure of \psad~is useful to obtain predictions for non-gravitational signatures of dark matter. In particular, the annihilation rate can be
written as an integral over all relative velocities in the limit of relative distance going to zero. In paper II of this series \citep{Zavala_13} we show
how this can be used to obtain predictions for annihilation signals for general velocity-dependent cross-sections. 

\section*{Acknowledgments}

We are grateful to the anonymous reviewer for useful suggestions and comments.
We thank the members of the Virgo consortium for access to the Aquarius simulation suite and our special thanks to
Volker Springel for providing access to SUBFIND and GADGET-3 whose routines on the computation of the 2PCF in
real space were used as a basis for the code in phase space we developed.
We also thank Simon White, Ed Bertschinger and Steen Hansen for useful suggestions.
JZ and NA are supported by the University of 
Waterloo and the Perimeter Institute for Theoretical Physics. 
Research at Perimeter Institute is supported by the Government of Canada 
through Industry Canada and by the Province of Ontario through the Ministry of Research and Innovation. JZ acknowledges financial support by a 
CITA National Fellowship. This work was made possible by the facilities of the Shared Hierarchical 
Academic Research Computing Network (SHARCNET:www.sharcnet.ca) and Compute/Calcul Canada.

\bibliography{lit}

\begin{thebibliography}{}

\bibitem[\protect\citeauthoryear{{Afshordi}, {Mohayaee} \&
  {Bertschinger}}{{Afshordi} et~al.}{2009}]{Afshordi_09}
{Afshordi} N.,  {Mohayaee} R.,    {Bertschinger} E.,  2009, \prd, 79, 083526

\bibitem[\protect\citeauthoryear{{Afshordi}, {Mohayaee} \&
  {Bertschinger}}{{Afshordi} et~al.}{2010}]{Afshordi_10}
{Afshordi} N.,  {Mohayaee} R.,    {Bertschinger} E.,  2010, \prd, 81, 101301

\bibitem[\protect\citeauthoryear{{Arkani-Hamed}, {Finkbeiner}, {Slatyer} \&
  {Weiner}}{{Arkani-Hamed} et~al.}{2009}]{Arkani_Hamed_09}
{Arkani-Hamed} N.,  {Finkbeiner} D.~P.,  {Slatyer} T.~R.,    {Weiner} N.,
  2009, \prd, 79, 015014

\bibitem[\protect\citeauthoryear{{Bertschinger}}{{Bertschinger}}{1985}]{Bertschinger_85}
{Bertschinger} E.,  1985, \apjs, 58, 39

\bibitem[\protect\citeauthoryear{{Boylan-Kolchin}, {Springel}, {White} \&
  {Jenkins}}{{Boylan-Kolchin} et~al.}{2010}]{BK_10}
{Boylan-Kolchin} M.,  {Springel} V.,  {White} S.~D.~M.,    {Jenkins} A.,  2010,
  \mnras, 406, 896

\bibitem[\protect\citeauthoryear{{Boylan-Kolchin}, {Springel}, {White},
  {Jenkins} \& {Lemson}}{{Boylan-Kolchin} et~al.}{2009}]{BK_09}
{Boylan-Kolchin} M.,  {Springel} V.,  {White} S.~D.~M.,  {Jenkins} A.,
  {Lemson} G.,  2009, \mnras, 398, 1150

\bibitem[\protect\citeauthoryear{{Davis} \& {Peebles}}{{Davis} \&
  {Peebles}}{1977}]{Davis_77}
{Davis} M.,  {Peebles} P.~J.~E.,  1977, \apjs, 34, 425

\bibitem[\protect\citeauthoryear{{Diemand} \& {Kuhlen}}{{Diemand} \&
  {Kuhlen}}{2008}]{Diemand_08}
{Diemand} J.,  {Kuhlen} M.,  2008, \apjl, 680, L25

\bibitem[\protect\citeauthoryear{{Gao}, {Frenk}, {Boylan-Kolchin}, {Jenkins},
  {Springel} \& {White}}{{Gao} et~al.}{2011}]{Gao_11}
{Gao} L.,  {Frenk} C.~S.,  {Boylan-Kolchin} M.,  {Jenkins} A.,  {Springel} V.,
    {White} S.~D.~M.,  2011, \mnras, 410, 2309

\bibitem[\protect\citeauthoryear{{Maciejewski}, {Vogelsberger}, {White} \&
  {Springel}}{{Maciejewski} et~al.}{2011}]{Maciejewski_11}
{Maciejewski} M.,  {Vogelsberger} M.,  {White} S.~D.~M.,    {Springel} V.,
  2011, \mnras, 415, 2475

\bibitem[\protect\citeauthoryear{{Navarro}, {Frenk} \& {White}}{{Navarro}
  et~al.}{1996}]{Navarro_96}
{Navarro} J.~F.,  {Frenk} C.~S.,    {White} S.~D.~M.,  1996, \apj, 462, 563

\bibitem[\protect\citeauthoryear{{Navarro}, {Frenk} \& {White}}{{Navarro}
  et~al.}{1997}]{Navarro_97}
{Navarro} J.~F.,  {Frenk} C.~S.,    {White} S.~D.~M.,  1997, \apj, 490, 493

\bibitem[\protect\citeauthoryear{{Navarro}, {Ludlow}, {Springel}, {Wang},
  {Vogelsberger}, {White}, {Jenkins}, {Frenk} \& {Helmi}}{{Navarro}
  et~al.}{2010}]{Navarro_10}
{Navarro} J.~F.,  {Ludlow} A.,  {Springel} V.,  {Wang} J.,  {Vogelsberger} M.,
  {White} S.~D.~M.,  {Jenkins} A.,  {Frenk} C.~S.,    {Helmi} A.,  2010,
  \mnras, 402, 21

\bibitem[\protect\citeauthoryear{{Springel}, {Wang}, {Vogelsberger}, {Ludlow},
  {Jenkins}, {Helmi}, {Navarro}, {Frenk} \& {White}}{{Springel}
  et~al.}{2008}]{Springel_08}
{Springel} V.,  {Wang} J.,  {Vogelsberger} M.,  {Ludlow} A.,  {Jenkins} A.,
  {Helmi} A.,  {Navarro} J.~F.,  {Frenk} C.~S.,    {White} S.~D.~M.,  2008,
  \mnras, 391, 1685

\bibitem[\protect\citeauthoryear{{Springel}, {White}, {Tormen} \&
  {Kauffmann}}{{Springel} et~al.}{2001}]{Springel_01}
{Springel} V.,  {White} S.~D.~M.,  {Tormen} G.,    {Kauffmann} G.,  2001,
  \mnras, 328, 726

\bibitem[\protect\citeauthoryear{{Taylor} \& {Navarro}}{{Taylor} \&
  {Navarro}}{2001}]{Taylor_01}
{Taylor} J.~E.,  {Navarro} J.~F.,  2001, \apj, 563, 483

\bibitem[\protect\citeauthoryear{{Vera-Ciro}, {Helmi}, {Starkenburg} \&
  {Breddels}}{{Vera-Ciro} et~al.}{2013}]{Vera-Ciro_13}
{Vera-Ciro} C.~A.,  {Helmi} A.,  {Starkenburg} E.,    {Breddels} M.~A.,  2013,
  \mnras, 428, 1696

\bibitem[\protect\citeauthoryear{{Vogelsberger}, {Helmi}, {Springel}, {White},
  {Wang}, {Frenk}, {Jenkins}, {Ludlow} \& {Navarro}}{{Vogelsberger}
  et~al.}{2009}]{Vogelsberger_09}
{Vogelsberger} M.,  {Helmi} A.,  {Springel} V.,  {White} S.~D.~M.,  {Wang} J.,
  {Frenk} C.~S.,  {Jenkins} A.,  {Ludlow} A.,    {Navarro} J.~F.,  2009,
  \mnras, 395, 797

\bibitem[\protect\citeauthoryear{{Vogelsberger} \& {White}}{{Vogelsberger} \&
  {White}}{2011}]{Vogelsberger_11}
{Vogelsberger} M.,  {White} S.~D.~M.,  2011, \mnras, 413, 1419

\bibitem[\protect\citeauthoryear{{Vogelsberger}, {White}, {Helmi} \&
  {Springel}}{{Vogelsberger} et~al.}{2008}]{Vogelsberger_08}
{Vogelsberger} M.,  {White} S.~D.~M.,  {Helmi} A.,    {Springel} V.,  2008,
  \mnras, 385, 236

\bibitem[\protect\citeauthoryear{{Zavala} \& {Afshordi}}{{Zavala} \&
  {Afshordi}}{2013}]{Zavala_13}
{Zavala} J.,  {Afshordi} N.,  2013, arXiv:1311.3296

\end{thebibliography}

\appendix

\section{Convergence tests}\label{conv}

\begin{figure*}
\begin{tabular}{|@{}l@{}|@{}l@{}|}
\includegraphics[height=8.0cm,width=8.0cm]{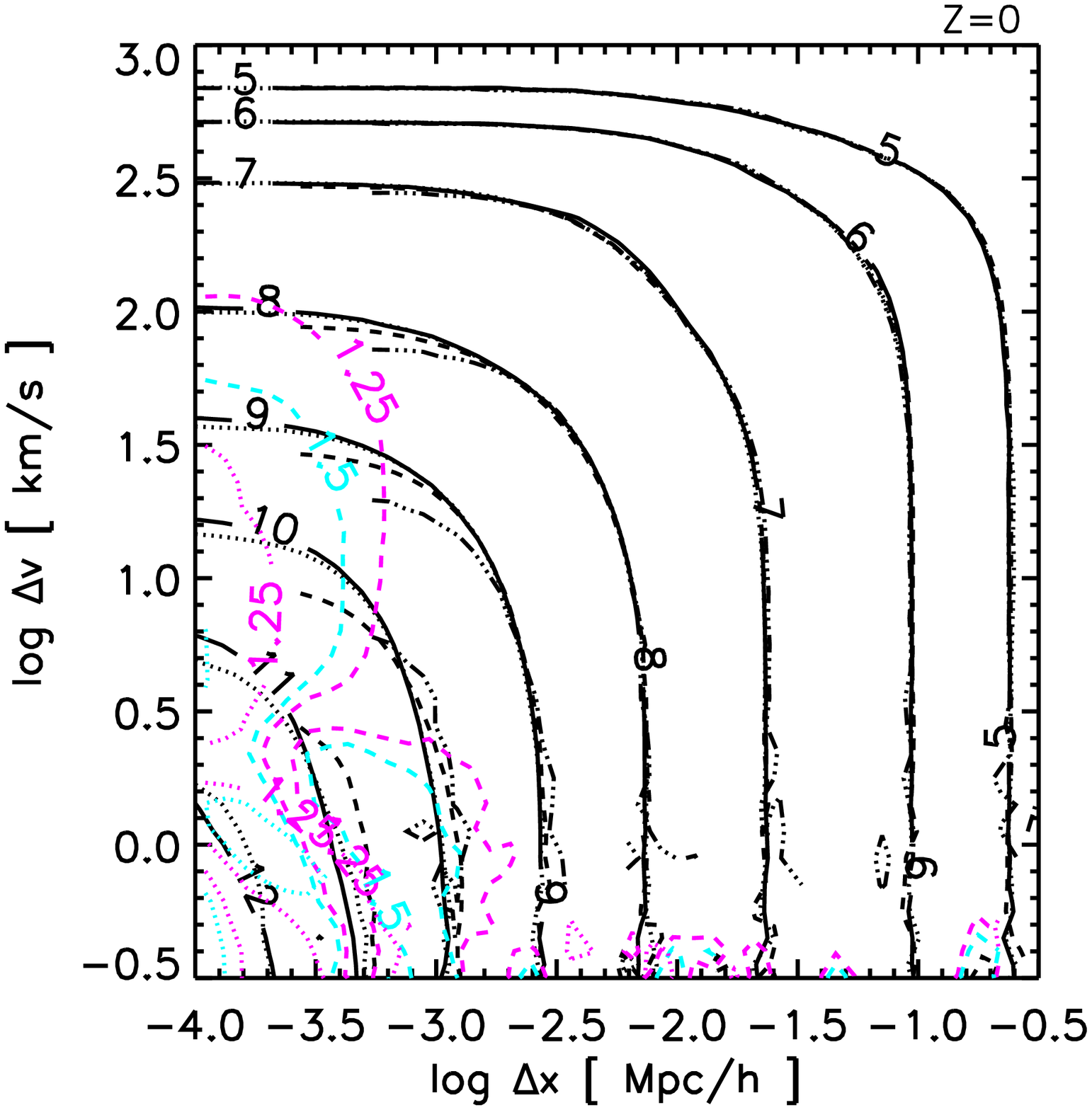} &
\includegraphics[height=8.0cm,width=8.0cm]{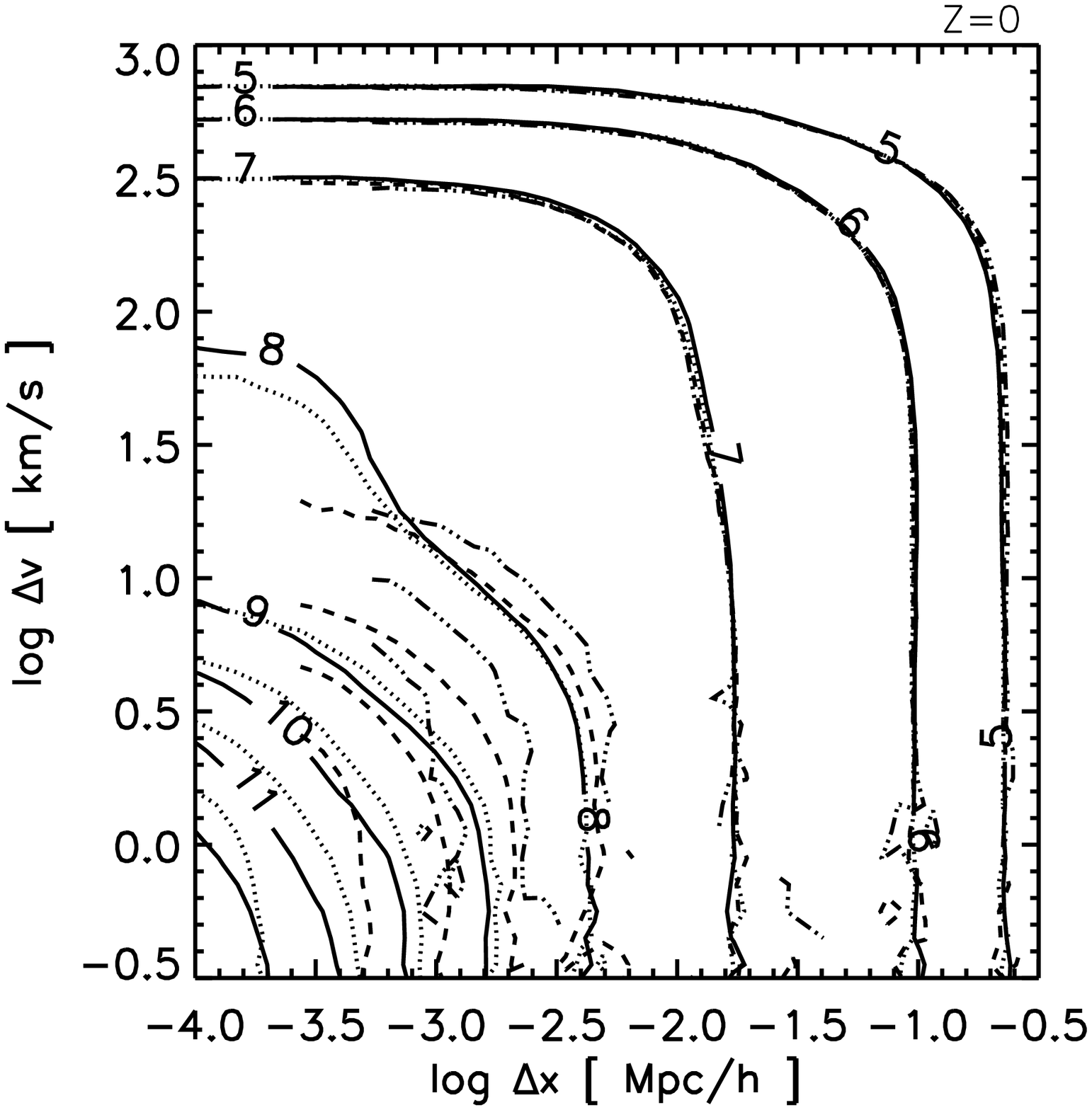} \\
\includegraphics[height=8.0cm,width=8.0cm]{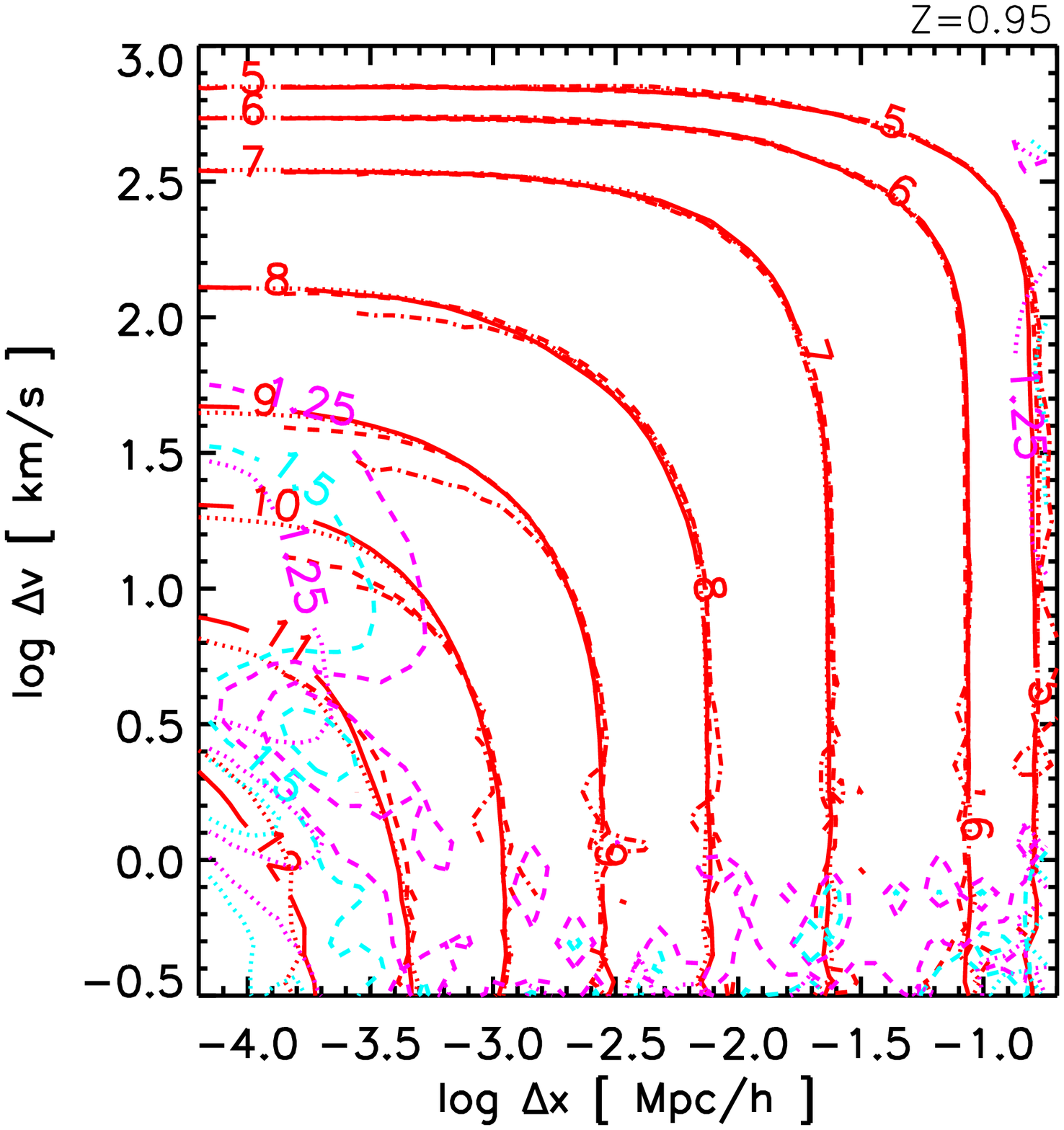} &
\includegraphics[height=8.0cm,width=8.0cm]{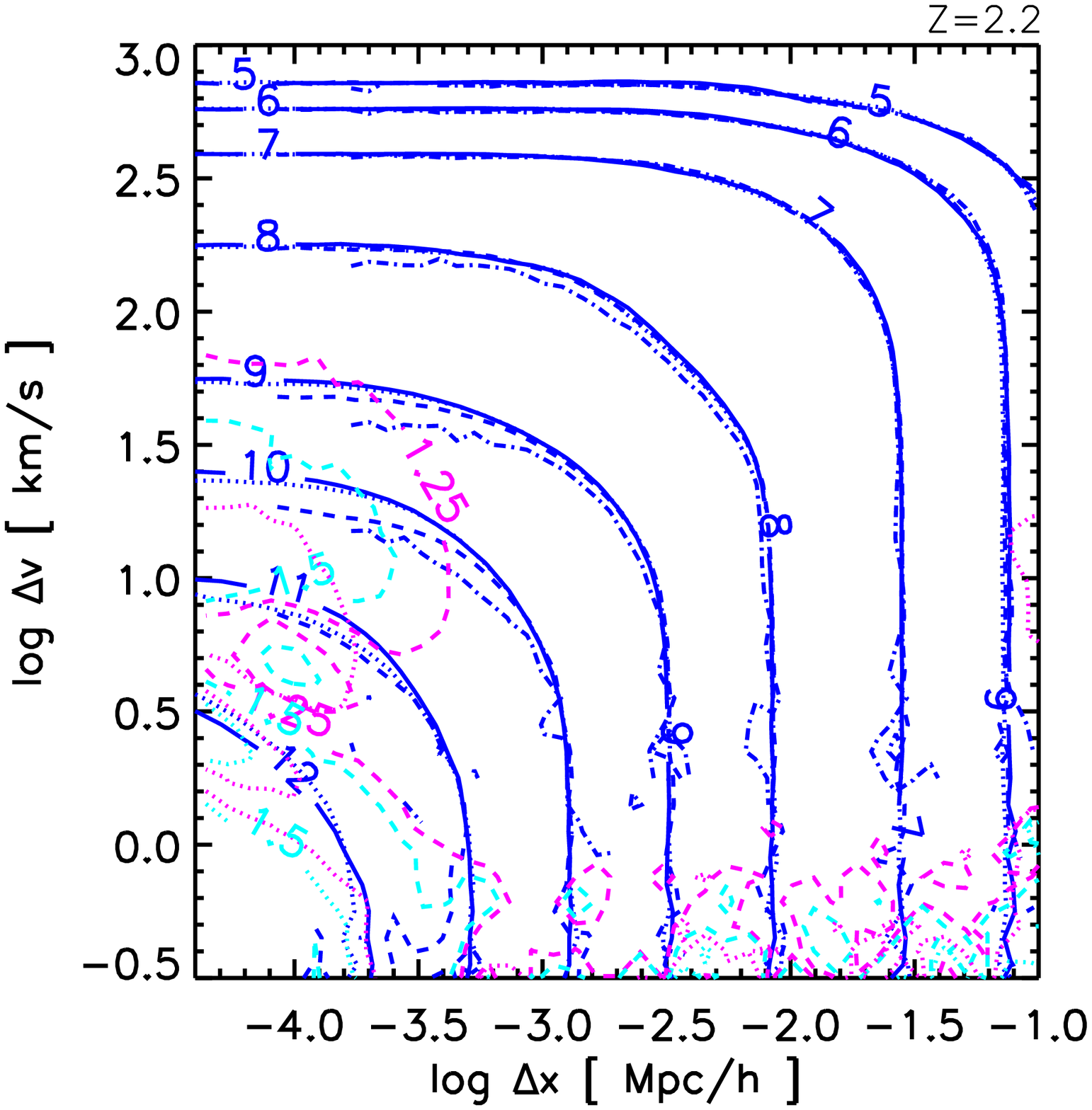} \\
\end{tabular}
\caption{Contours of the logarithm of the particle phase space average density (\psad) for the Aquarius A-halo  
at different resolutions levels: 2(solid), 3 (dotted), 4 (dashed) 
and 5 (dotted-dashed). {\it Upper panels}: At $z=0$ for the main halo and its subhaloes (left) and for the main
halo only (right). {\it Bottom panels}: $z=0.95$ (red, left), $z=2.2$ (blue, right) for the main halo and its subhaloes.
The magenta and cyan contours mark the regions where the value of the coarse-grained
phase space density between two consecutive resolution levels (excluding level 5 since it is dominated by noise at small separations
in velocity) is a factor of 1.25 and 1.5, respectively: 3 and 4 (dashed), 2 and 3 (dotted).}
\label{Fig_host_vs_all} 
\end{figure*}

Fig.~\ref{Fig_host_vs_all} shows the particle phase space average density (\psad) computed with our algorithm for the Aq-A halo at
different resolutions, from level 5 (dashed-dotted line) to level 2 (solid line). The upper-left ($z=0$), lower-left ($z=0.95$) and
bottom-right ($z=2.2$) panels show the case where particles in the main halo and its subhaloes contribute to the
average, while the upper-right panel ($z=0$) considers only particles within the main halo. As we mentioned in Section \ref{smooth}, 
the main host dominates \psad~at large $(\Delta x,\Delta v)$, while at smaller scales, the contribution
of the smooth halo is subdominant and it is of course resolution dependent, since simulation particles that are part of
the smooth component at low resolutions can form bound subhaloes at higher resolutions. This is clearly shown in 
in the right panel of Fig.~\ref{Fig_host_vs_all}. At large $(\Delta x,\Delta v)$, the value has clearly converged, but at small separations 
in phase space, the smooth component is ever decreasing with increasing resolution. On the contrary, \psad~for both the smooth and subhalo
components (upper-left panel) is roughly convergent until the value of the separation that can be resolved for a given resolution. 
By increasing the resolution, the values that \psad~can take are larger and larger reaching smaller and smaller separations in phase space. 
We also note that sampling errors (a random number of particles $N_s$ is used to estimate \psad, see Section \ref{l_estimator})
are negligible. This can be seen by noticing that although we keep the samples of particles fixed to a few millions for the simulations across 
different resolutions, these samples are of course chosen at random in {\it each} simulation. If sampling errors were an issue, then convergence 
at large separations in phase space would be much poorer. We have nevertheless tested this explicitly taking larger samples and found no significant
difference in our results.

The spatial resolution of each simulation is roughly characterized by the scale at which the gravitation law is ``softened'' $\sim2.8\epsilon$, where
$\epsilon$ is the Plummer-equivalent gravitational softening length. Since we are dealing with convergence in phase space we determine instead 
the region in phase space where \psad~has converged by over-plotting in Fig.~\ref{Fig_host_vs_all} contours of constant ratio of \psad~between
consecutive resolution levels, e.g. a ratio of 1.25 between resolution levels 3 and 2 is given by a magenta dotted line. We only do so for two values
of this ratio, 1.25 and 1.5. By looking at the contours for a fixed value of this ratio at increasing resolution we can confidently establish that
for level 2, resolution is not an issue for most of the region in phase space explored in this paper.

\begin{figure*}
\begin{tabular}{|@{}l@{}|@{}l@{}|}
\includegraphics[height=8.0cm,width=8.0cm]{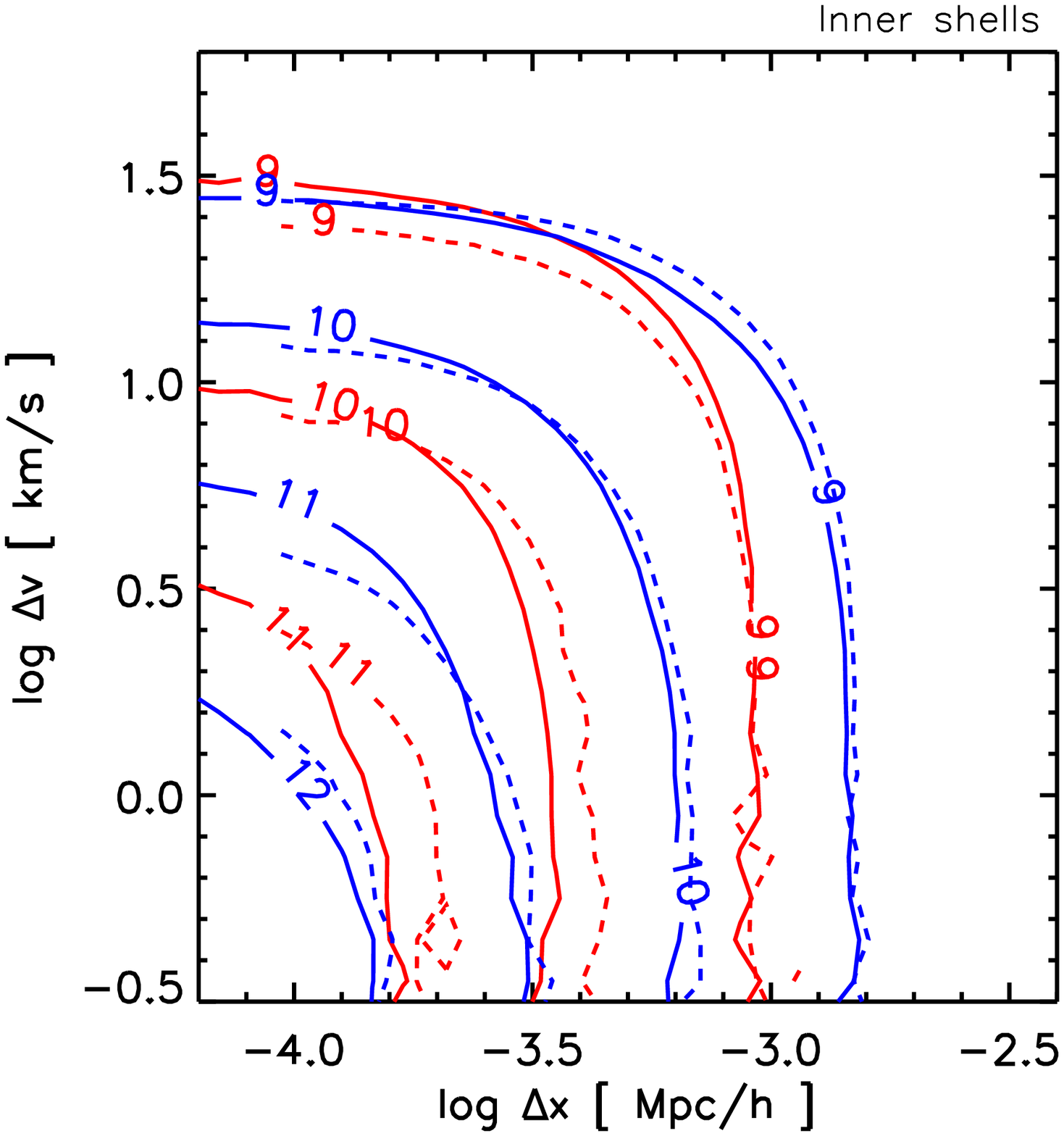} &
\includegraphics[height=8.0cm,width=8.0cm]{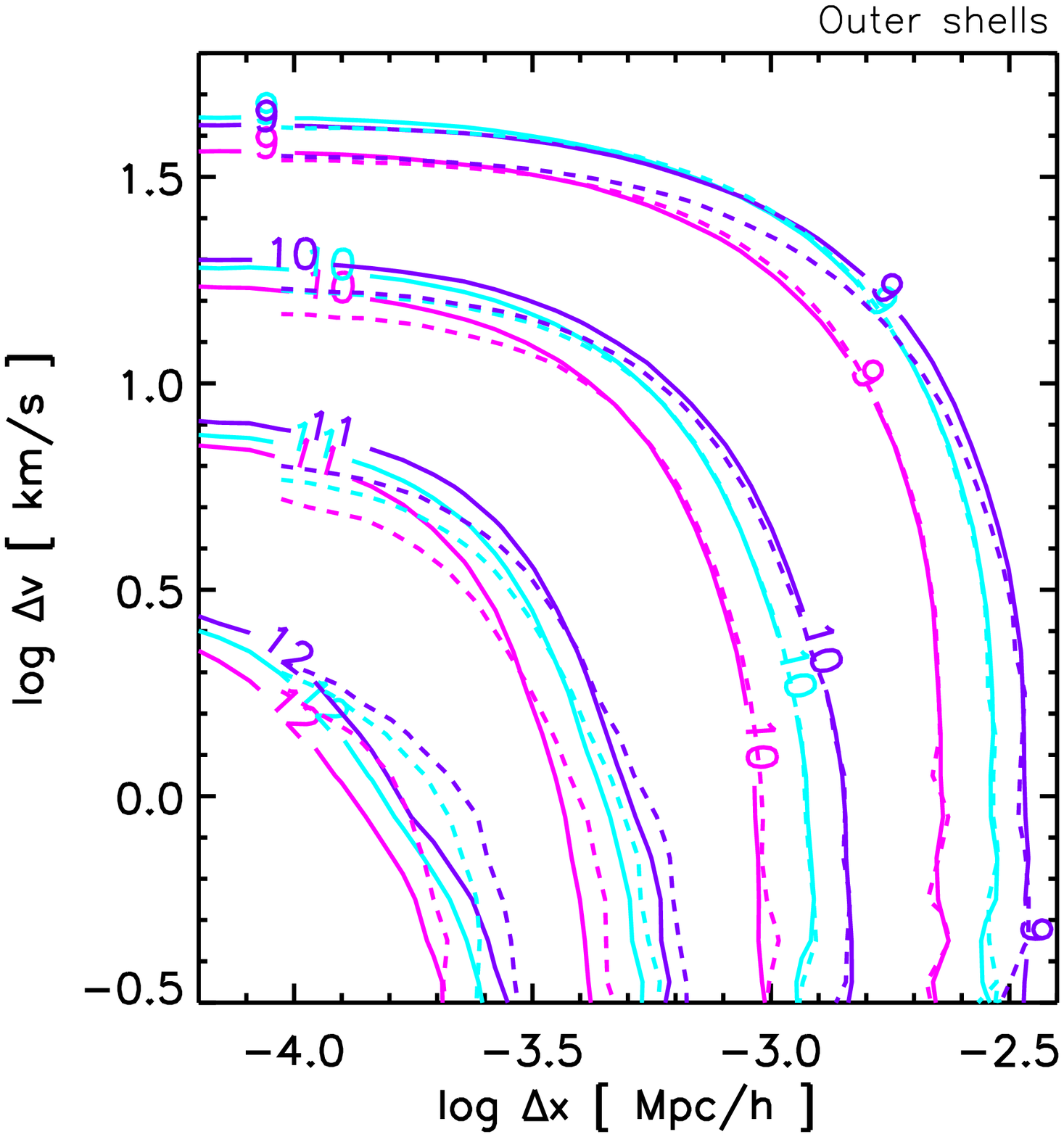} \\
\end{tabular}
\caption{Contours of the logarithm of the particle phase space average density (\psad) for the Aquarius A-halo at $z=0$ for two different
resolution levels:  Aq-A-3 (dashed) and Aq-A-2 (solid), at different radial shells from the halo centre: 
$0<r/r_{200}<0.2$ (red, left), $0.2<r/r_{200}<0.4$ (blue, left), $0.4<r/r_{200}<0.6$ (magenta, right), $0.6<r/r_{200}<0.8$ 
(cyan, right), and $0.8<r/r_{200}<1.0$ (violet, right).} 
\label{Fig_convergence_radial} 
\end{figure*}

Fig.~\ref{Fig_convergence_radial} shows the convergence for \psad~averaged over different volumes corresponding to
concentric shells of thickness $0.2r_{200}$ from the centre of the halo. All particles within a given shell
are considered for two resolution levels at $z=0$: level 3 (dashed) and level 2 (solid). Although the results seem to converge
across resolutions, they do so with larger deviations compared to when the average is taken within larger volumes. This
is likely related to the fluctuations in subhalo abundance as a function of radius (see e.g. the middle panel of Fig. 12
of \citealt{Springel_08}).

\end{document}